\documentclass{aa}  

\usepackage{graphicx}
\usepackage[varg]{txfonts}

\newcommand{\tastar}{\hbox {T$_{\rm A}^*$ }}
\newcommand{\tastarp}{\hbox {T$_{\rm A}^*$}}

\newcommand{\twco}{{\hbox {\ensuremath{\mathrm{^{12}CO}} }}}
\newcommand{\twcop}{{\hbox {\ensuremath{\mathrm{^{12}CO}}}}}
\newcommand{\ceo}{{\hbox {\ensuremath{\mathrm{C^{18}O}} }}}
\newcommand{\ceop}{{\hbox {\ensuremath{\mathrm{C^{18}O}}}}}

\newcommand{\thco}{{\hbox {\ensuremath{\mathrm{^{13}CO}} }}}
\newcommand{\thcop}{{\hbox {\ensuremath{\mathrm{^{13}CO}}}}}

\newcommand{\kmps}{\ensuremath{\mathrm{km\,s^{-1}}}}
\newcommand{\percubcm}{\ensuremath{\mathrm{cm^{-3}}}}
\newcommand{\persqcm}{\ensuremath{\mathrm{cm^{-2}}}}
\newcommand{\Msun}{\ensuremath{\mathrm{M}_\odot}}
\newcommand{\Msunp}{\ensuremath{\mathrm{M}_\odot}.}

\newcommand{\Lsun}{\ensuremath{\mathrm{L}_\odot}}
\newcommand{\kkms}{\ensuremath{\mathrm{K\,km\,s^{-1}}}}
\newcommand{\hour}{\ensuremath{^\mathrm{h}}}
\newcommand{\minute}{\ensuremath{^\mathrm{m}}}

\newcommand{\Tmb}{\ensuremath{\mathrm{T}_\mathrm{MB}}}

\newcommand{\Htwo}{\ensuremath{\mathrm{H}_2}}
\newcommand{\Halpha}{\ensuremath{\mathrm{H}\alpha}} 
\newcommand{\mum}{\mu {\rm m}}
\def\Ks{\hbox{$K$s}}
\def\K{\hbox{$K$}}

\def\J{\hbox{$J$}}
\def\H{\hbox{$H$}}
\newcommand{\umag}{{$^{m}$}}
 
\def\UBV{\hbox{$U\!B\!V$}}

\def\yzjhk{\hbox{$Y\!Z\!J\!H\!K$}} 
 
\def\jhks{\hbox{$J\!H\!K$s}}              

\begin{document}
   \title{The Bok Globule BHR\,160: structure and star formation 
     \thanks{Based on observations made with ESO Telescopes at the La
       Silla or Paranal Observatories.}\thanks {Reduced molecular line spectra are  available in electronic form
at the CDS via anonymous ftp to cdsarc.u-strasbg.fr (130.79.128.5)
or via http://cdsweb.u-strasbg.fr/cgi-bin/qcat?J/A+A/}}

 \author{L. K. Haikala\inst{1}
          \and
          Bo Reipurth\inst{2}
          }

   \institute{Universidad de Atacama, Copayapu 485, Copiapo, Chile\\
              \email{lkhaikala@gmail.com}
         \and
              Institute for Astronomy, University of Hawaii at Manoa, 
640 N. Aohoku Place, Hilo, HI 96720, USA\\
              \email{reipurth@ifa.hawaii.edu}
 }

    \date{}
 
  \abstract
   {BHR\,160 is a virtually unstudied cometary globule within the Sco~OB4
association in Scorpius at a distance of 1600~pc. It is part of a
system of cometary clouds which face the luminous O star HD155806.
BHR\,160 is special because it has an intense bright rim. 
    }
   {We attempt to derive physical parameters for BHR\,160 and to understand
its structure and the origin of its peculiar bright rim. }
   {BHR\,160 was mapped in the \twcop, \thco
   and \ceo (2--1) and (1--0) and CS (3--2) and (2--1) lines. These
   data, augmented with stellar photometry derived from the ESO VVV survey,
   were used to derive the mass and distribution of molecular material
   in BHR\,160 and its surroundings. Archival mid-infrared data from the WISE
   satellite was used to find IR excess stars in the globule and
   its neighbourhood.}
   {An elongated 1\arcmin\ by 0\farcm6
   core lies adjacent to the globule bright rim. \twco emission covers
   the whole globule, but the \thcop, \ceo and CS emission is more
   concentrated to the core.  The \twco line profiles indicate the
   presence of outflowing material near the core, but the spatial
   resolution of the mm data is not sufficient for a detailed spatial
   analysis. The BHR\,160 mass estimated from the \ceo mapping is {
     100$\pm$50\,\Msun\ (d/1.6\,kpc)$^2$ where d is the distance
     to the globule. Approximately 70\% of the mass lies in the dense
     core.  The total mass of molecular gas in the direction of
     BHR\,160 is 210$\pm$80\,(d/1.6\,kpc)$^2$\,\Msun\ when estimated from
     the more extended VVV NIR photometry.}  We argue that the bright
   rim of BHR\,160 is produced by a close-by early B-type star,
   HD~319648, that was likely recently born in the globule.  This star
   is likely to have triggered the formation of a source, IRS~1, that
   is embedded within the core of the globule and detected only
     in \Ks\ and by WISE and IRAS.  }
     {}
     
   \keywords{Stars: formation -- Stars:pre-main-sequence --
    ISM:individual, BHR\,160 -- ISM:dust, extinction
               }

   \titlerunning{Bright rimmed globule BHR\,160 }
   \maketitle

\authorrunning{L.K. Haikala}
\section{Introduction} \label{sect:introduction}

Bok globules are small compact clouds with typical dimensions of
approximately 0.15 to 0.8~pc \citep{Bok1977, reipurth2008}. { Many of
these are cloud cores} which have been exposed from the interior of
large molecular clouds when the more tenuous material has been swept
away by the formation of nearby OB stars \citep{reipurth1983}.
Cometary globules are a subset of Bok globules in a transition phase,
still showing the windswept appearance of their formation. The
compression that the globules suffer as they are being excavated, in
many cases leads to the formation of stars, so cometary globules are
frequently containing young stars
\citep[for example][]{haikalareipurth2010,haikalaetal2010}.

In a survey for globules in the Galactic plane, we have come across a
remarkable yet virtually unexplored cometary globule. It is listed as
object \object {353.3+2.4} in the \citet{hartleyetal1986} list of southern dark
clouds. \citet{bourkeetal1995a, bourkeetal1995b} surveyed the optical
appearance of the opaque \citet{hartleyetal1986} clouds smaller than
10$'$ in diameter and searched for ammonia (1,1) emission in this
selection. Globule 353.3+2.4 is object 160 in the
\citet{bourkeetal1995a} list and will be identified as \object {BHR\,160} in the
following.

The globule is clearly cometary, but the striking aspect of it is its
bright rim (Figure~1). The brightest part of this rim has a width of
$\sim$2$'$ (0.9~pc) with fainter extensions on both sides.  We have
obtained a poor, noisy red spectrum { at the ESO 3.6m telescope} of the
bright rim which shows a strong H$\alpha$ line and much weaker lines
of [SII] $\lambda$~6717/6731 lines, as expected in photoionized gas.
None of the other cometary clouds in the region show similar bright
rims, which leads to the suspicion that a bright star $\sim$30" away
from the midpoint of the bright rim is the source of UV radiation.
This star is \object {HD 319648}, an early B star.

The general region of BHR\,160 contains a number of cometary globules
and cometary-shaped clouds, which all { face towards another, more
distant,} bright star, \object {HD 155806} (= HR~6397 = V1075~Sco), see Figure~2.
This is an O7.5Ve star \citep{walborn1973}, and the hottest known
Galactic Oe star \citep{fullertonetal2011}.  HD~155806 is the most
luminous star in the little-studied \object {Sco OB4} association
\citep{the1961, roslund1966}.  Within a radius of one degree around
HD~155806, \citet{the1961} found 40 OB stars and 85 A-type stars, for
which he adopted a distance of $\sim$1400~pc. In a follow-up study,
\citet{roslund1966} found that the Sco OB4 association extends towards
the south, with its main concentration in the young H~II region NGC
6334.  \citet{persitapia2008} have determined the distance to NGC~6334
to be 1.61$\pm$0.08~kpc.  In the following we adopt a distance of
1.6~kpc for BHR\,160.

In this paper, we present detailed millimetre-wavelength
multi-transition observations of the virtually { unstudied} globule
BHR\,160, and we use the data to determine the mass and structure of
the globule. The observations are augmented with archival near- and
mid-infrared data which reveal that star formation is ongoing within
the globule.

\section{Observations and data reduction} \label{sect:observations}

The single BHR\,160 ammonia (1,1) spectrum in \citet {bourkeetal1995b}
was a non-detection. This is probably explained by the 1\farcm4 HPBW
of the Parkes telescope used in the survey. The beam covers
practically the whole globule and a dense, small-size core is highly
beam diluted. Spectral line observations with better spatial
resolution and molecular species and transitions tracing also the more
diffuse material in BHR\,160 are called for.  The CO molecule and its
isotopologues trace gas column density, and especially the \twco
molecule is excited already at moderate densities. As the abundance of
\thco and \ceo is only a fraction of that of \twcop, their optical
depths and consequently -- because of less radiative trapping  -- their
effective critical densities are higher. The CS molecule is a good
tracer for high-density gas.  This makes the CO (and isotopologues)
(2--1) and (1--0) and the CS (3--2) and (2--1) transitions convenient
tools when studying the basic properties of a moderately dense globule
core and its less dense envelope.

\begin{figure}
 \centering
 \includegraphics [width=8.8cm]{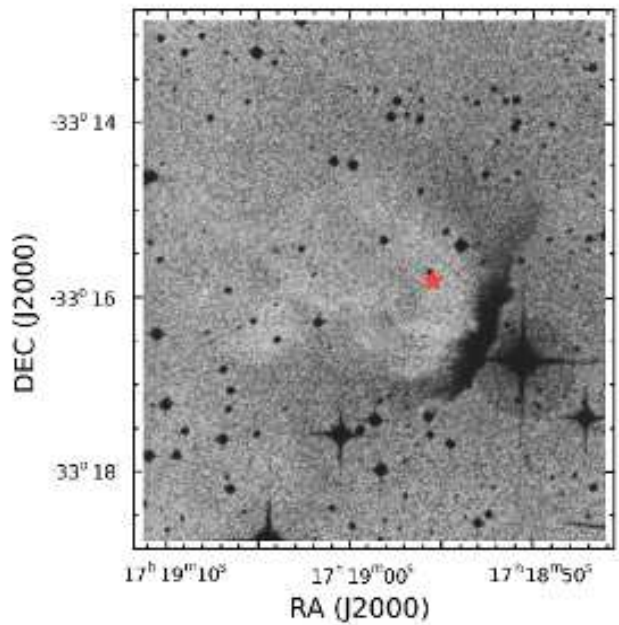}
 \caption{Close-up of the BHR\,160 globule from an ESO-R sky survey
   plate.  The bright rim is striking in appearance. The bright star
   in front of the bright rim is the early B-type star HD~319648. {
     The red asterisk marks the location of the embedded source IRS~1,
     see Fig.~\ref{fig:WISE} and the text for details. The region
     between IRS~1 and the bright rim is here called ‘the core’.}}
 \label{fig:ESO-R}
 \end{figure}

\begin{figure}
 \centering
\includegraphics [width=8.8cm]{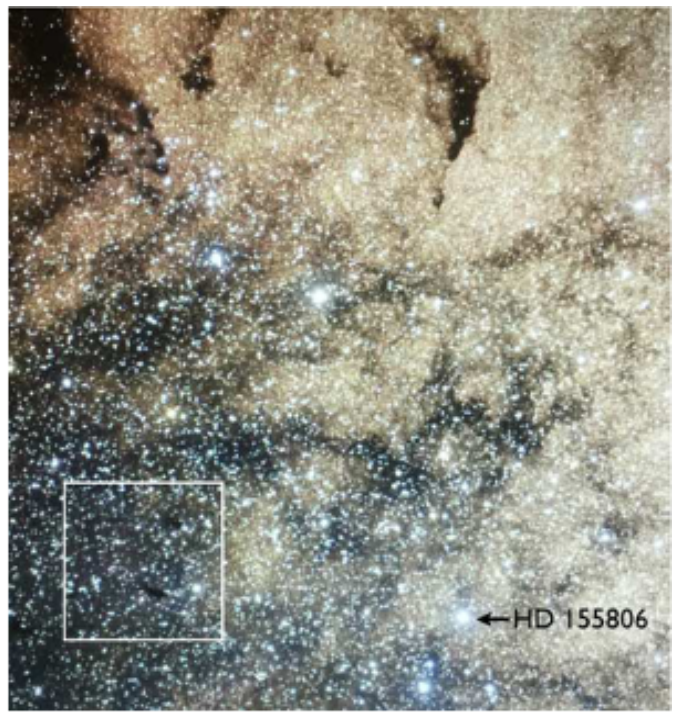}
    \caption{Northern part of the Sco OB4 association includes the
      luminous O7.5V star HD~155806, towards which BHR\,160 and other
      cometary-shaped clouds point. The area of Fig. \ref{fig:AAO}, which
      includes BHR\,160, is indicated by a rectangle.  Image courtesy
      S. Guisard/ESO}
 \label{fig:ScoOB4}
 \end{figure}

The molecular line observations were obtained in September and October
1999 with the Swedish-ESO-Submillimetre-Telescope, SEST, at the La
Silla observatory, Chile. The SEST 3 and 2~mm~(SESIS) and 3 and
1~mm~(IRAM) dual SiS SSB receivers were used. The IRAM receiver was
used except when observing the CS lines, for which the SESIS receiver
was used. The SEST high-resolution 2000 channel acousto-optical
spectrometer (bandwidth 86\,MHz, channel width 43\,kHz) was split into
two to measure two receivers simultaneously. At the observed
wavelengths, 3~mm, 2~mm and 1~mm, the 43\,kHz channel width corresponds to
\hbox{$\sim0.12$~\kmps}, \hbox{$\sim0.08$~\kmps}\ and
\hbox{$\sim0.06$~\kmps}, respectively.

An approximately 2\farcm5 by 2\farcm5 area was observed with
20\arcsec\ spacing and a tilt of 60\degr. The map was centred at
$17\hour 18\minute 53\fs 1, -33\degr 16\arcmin 25\arcsec $~(J2000).
The CO\,(2--1) and (1--0) and the CS\,(3--2) and (2-1) transitions
were observed simultaneously.
  Frequency switching observing mode was used. The switch was 5\,MHz
  for the 1~mm and 15\,MHz for 2 and 3~mm observations. A second-order
  baseline was subtracted from the spectra after folding.  Calibration
  was achieved by the chopper wheel method. All the line temperatures
  in this paper, unless specially noted, are in the units of \tastarp,
  that is corrected to outside of the atmosphere but not for beam
  coupling.  Typical values for the effective SSB system temperatures
  outside the atmosphere were around 200~K except for the \twco (1--0)
  line for which it was 350~K. The RMS of the spectra for one minute
  integration after folding was around 0.1~K except for CO (1--0)
  (0.15~K). Pointing accuracy is estimated to be better than 5\arcsec.

The observed molecular transitions, their frequencies, SEST half power
beam width, HPBW, and the telescope main beam efficiency, $\eta_{\rm
mb}$, at these frequencies are given in Table \ref{table:obs}.  For
the CO observations only the \twco  is listed.  At a distance of 1.6\,kpc
the SEST HPBW at 230 GHz corresponds to 0.19\,pc.  
\begin{table}
  \caption[]{Observed lines and telescope parameters}
   \label{table:obs}
\begin{tabular}{lccc}
\hline
\hline
Line &  $\nu$   & HPBW & $\eta_{\rm mb}$ \\
     & (GHz) & ($\arcsec$)&  \\
\hline
CS (2--1)         & 97.271   & 54& 0.70     \\
CO (1--0)       & 115.271  & 47&  0.70\\
CS (3--2)         & 145.904  & 34&  0.66\\
CO (2--1)       &  230.538 & 24&  0.50 \\
\hline
\end{tabular}                                          

\end{table}

\section{Results} \label{sect:results}

\subsection{Optical and infrared archival data} \label{sect:archive}

Survey data of the BHR\,160 region can be found in the AAO/UKST
\Halpha\ survey \citep{parkeretal2005}, ESO VISTA variables in the
V\'ia L\'actea \citep[VVV,][]{saitoetal2012} survey, Wide-Field
Infrared Survey Explorer \citep [WISE,][]{wrightetal2010} and IRAS
public data bases. These data { were} used to evaluate the
distribution of interstellar matter and properties of individual
stellar sources in the region.

\begin{figure}
 \centering
 \includegraphics [width=8.8cm]{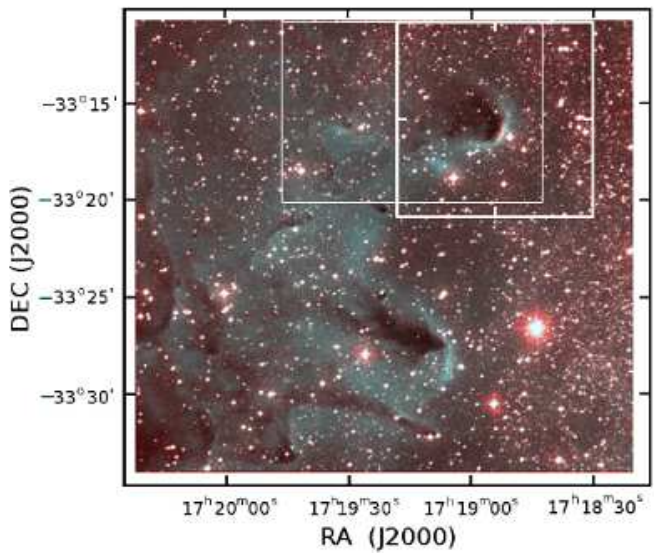}
 \caption{ { AAO/UKST survey image of BHR\,160 combining an
     \Halpha\ image (blue) and a short red broadband image (red).
     BHR\,160 is located in the upper right corner of the image, and
     the larger cloud to the south of NHR~160 is known as BHR\,159.
     The rectangles indicate the location of Figs. \ref{fig:WISE} and
     \ref {figure:BR1VVV}.}}
 \label{fig:AAO}
 \end{figure}

The colour coded image { from the AAO/UKST \Halpha\ survey (\Halpha\
  plus [N{\small II}] coded in blue) and short-exposure red-band
  (5900-6900\AA) image} (in red) shown in Fig. \ref{fig:AAO} reveals
that BHR\,160 is in fact a part of a receding wall of an (extinct) HII
region. One large cometary-shaped object { (\object {BHR\,159})} and several smaller
globules lie to the South-East.  The Western edge of BHR\,160 emits
strongly in \Halpha\ but the UKIDSS Y and Z images indicate that a
part of the surface brightness is also reflected light. The source of
the excitation is possibly the nearby B star HD\,319648
located very close to the globule.

\begin{figure}
 \centering
 \includegraphics [width=8.8cm]{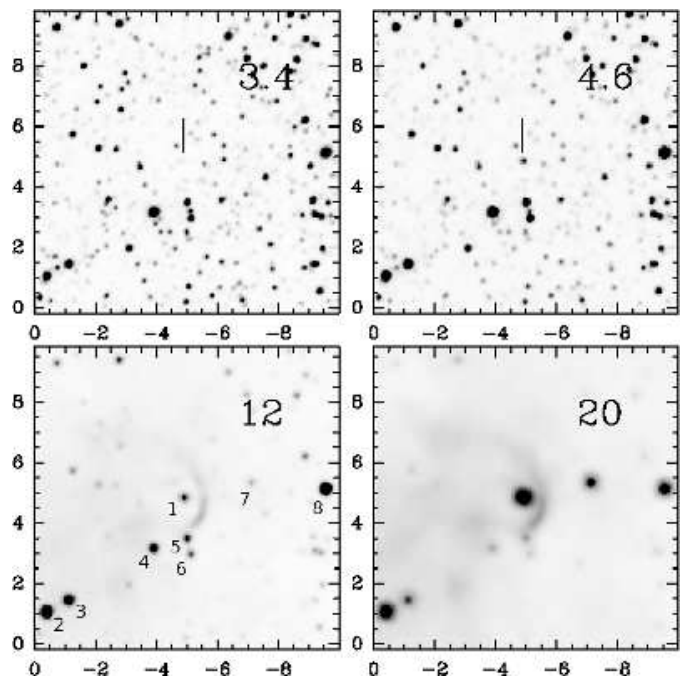}
 \caption{WISE images at 3.4, 4.6, 12 and 20 $\mum$. The location of
   BHR\,160\,IRS1 is indicated with a line in the two upper panels and
   with 1 in the lower left panel. Besides BHR\,160\,IRS1 seven
   further stellar sources emitting strongly at 12\,$\mum$ are marked
   in the latter panel.  The unit of the axes is arcminutes.}
 \label{fig:WISE}
 \end{figure}

 WISE imaged BHR\,160 at 3.4, 4.6, 12 and 20\,$\mum$ (Fig.
 \ref{fig:WISE}). Only stars are visible in the direction of the
 globule at the shortest wavelengths 3.4 and 4.6\,$\mum$. A star in
 the centre of the globule (\object{WISE J171855.54-331554.3}), which
 is faint at 3.4\,$\mum$\ (10\fm97), becomes brighter at 4.6\,$\mum$
 and continues brightening at 12 and 20\,$\mum$. This WISE source
 will be referred to as BHR\,160\,IRS1
 in the following. Besides BHR\,160\,IRS1 seven further stellar
 sources emitting strongly at 12\,$\mum$ are marked in the Figure. The
 globule bright rim is visible at the two longest wavelengths. Besides
 the bright rim, fainter surface brightness is also visible along the
 body of the globule. An IRAS point source \object{IRAS 17156-3312}
 lies 15\arcsec\ West of the WISE source. It has a good quality
 detection at 60 and 100\,$\mum$ bands, 15.3 and 71.8 Jy,
 respectively. In the HIRES processed \citep{aumanetal1990} IRAS
 images, the 60\,$\mum$ emission maximum coincides with the WISE source,
 but the 100\,$\mum$ emission maximum is shifted towards the bright
 rim. The spatial resolution of the IRAS 100\,$\mum$ observations,
 even when HIRES-processed, is not sufficient to resolve the small scale
 structure in the globule. Therefore it is likely that part of the
 observed 100\,$\mum$ emission comes from heated dust associated with
 the globule.

\begin{figure}
 \centering
 \includegraphics [width=8.8cm]{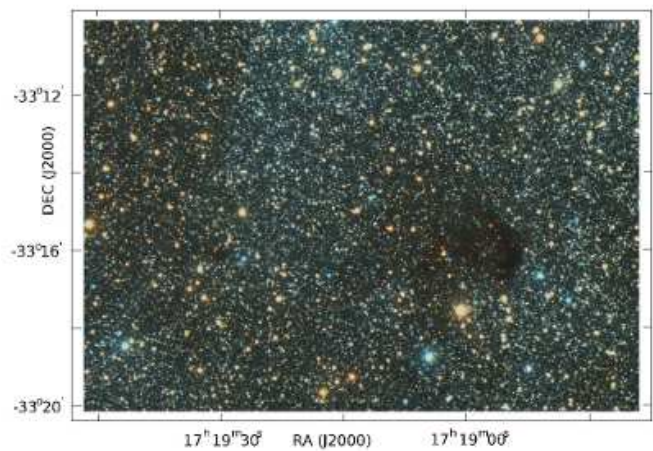}
 \caption{False colour VVV  \yzjhk\ image.}
 \label{figure:BR1VVV}
 \end{figure}

 In the \Halpha\ and the short-exposure red images
 (Fig.~\ref{fig:AAO}), optical extinction is high in an arc { to the
   east of (behind)} the bright rim, and a less extincted area
 continues to NE. This is in agreement with the ESO VISTA VVV survey
 \yzjhk \ false colour image shown in Fig. \ref{figure:BR1VVV}.
BHR\,160\,IRS\,1 is not situated inside the dense elongated core but at its
edge on the opposite side of the bright rim. A faint, slightly
extended non-stellar red object detected only in the \Ks\ band
coincides with BHR\,160\,IRS1.  The region of high extinction between
the bright rim and BHR\,160\,IRS1 will be called  ‘the core’ in the
following. No stars can be seen in the very centre of the core in Fig.
\ref{figure:BR1VVV}.

\subsection{Molecular line observations} \label{sect:lines}

\begin{figure}
 \centering
 \includegraphics [width=8.8cm]{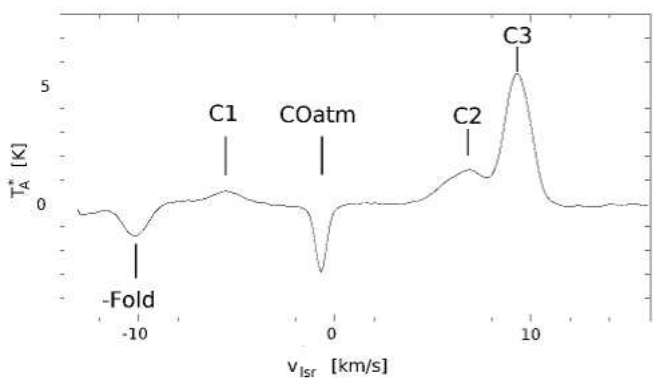}
    \caption{Folded \twco (2--1) line { (average of spectra around
        the centre of the map)}.  The three line components, the
      atmospheric line and the negative component due to the folding
      process are indicated.}
 \label{fig:co2-1line}
 \end{figure}

Three velocity components are seen in the \twco spectrum (Fig.
\ref{fig:co2-1line}) observed in the direction of BHR\,160. The
components will be referred to as C1, C2 and C3 as marked in the
Figure. Two negative line components are seen in this spectrum. The
negative line component at $\sim$--10\,\kmps \ is produced in the
frequency switching technique by component C3.  The negative signal at
$\sim$--1\,\kmps \ is due to atmospheric CO.  The strongest component,
C3, is detected in all \twco and \thco spectra and its emission is
centred on BHR\,160. The C2 \twco emission is distributed in a ridge
along the globule centre axis { (the line from SW to NE through the
  centre of the CO map).}   { Emission from the C1 component is detected in
  the SE edge of the CO map and is visible only in the \twco line. }

\begin{figure}
 \centering
  \includegraphics [width=8.8cm]{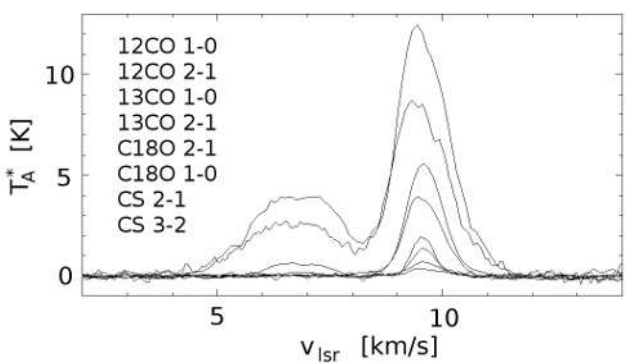}
    \caption{Observed CO and CS lines at the map off-set position
      +17\arcsec,10\arcsec.  Only the two strongest line components,
      C2 and C3, are shown { The lines are identified on the left
        in the order of the intensity of the C3 component.} }
 \label{fig:sample_spectra}
 \end{figure}

\begin{figure*}
 \centering
 \includegraphics [width=18cm]{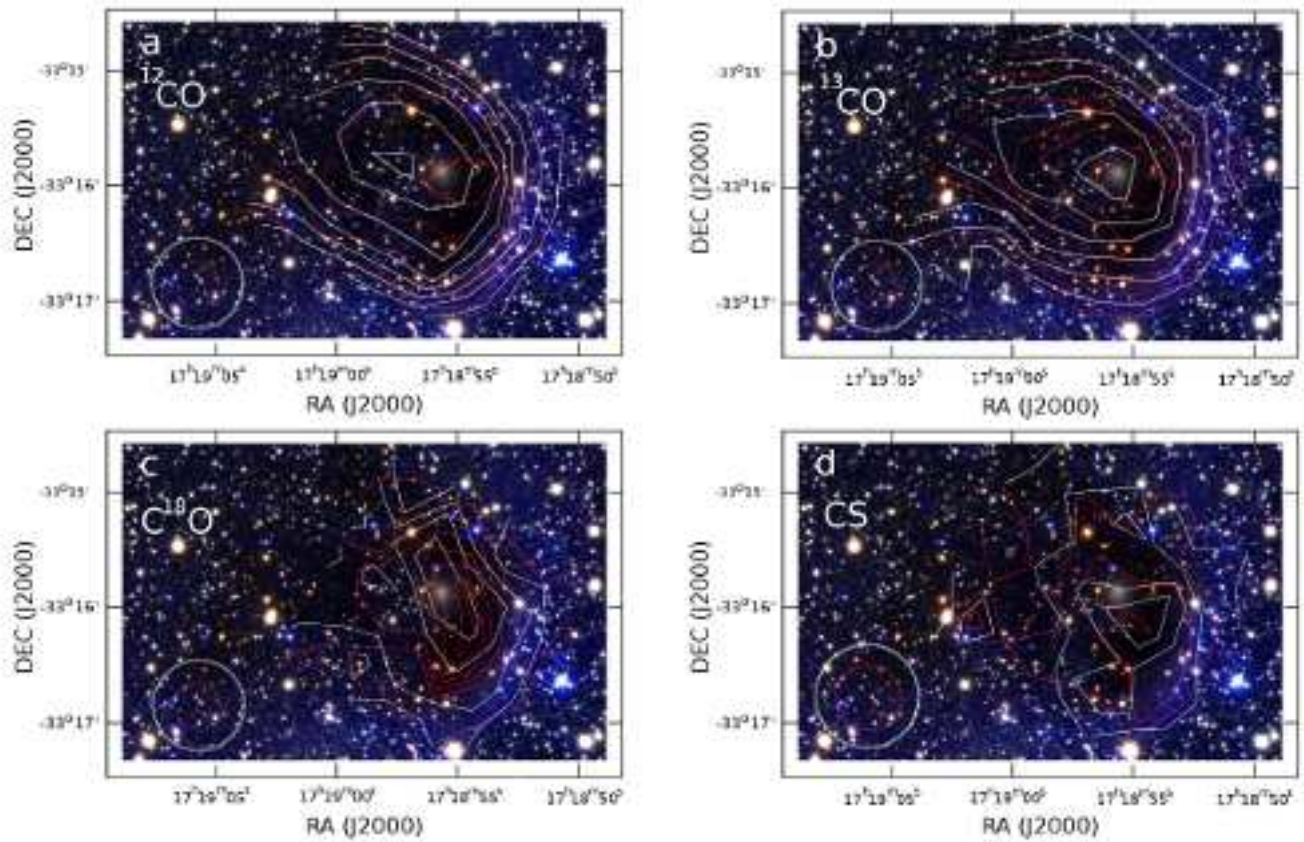}
    \caption{Contours of the line integrals of the observed molecular
      transitions { (only component C3)} overlayed on a VVV,
      \Halpha\ and WISE { false colour image.} The respective beam
      sizes for each molecule are shown in the lower left of each
      panel.  { a} \twco (1--0) (in white) and (2--1) (in red)
      contours. { Lowest contour level and increment in \tastar
        scale is 2\,\kkms .} { b} As a but for \thcop. { Lowest
        contour level and increment are 1\,\kkms\ for \thco (1--0) and
        0.5\,\kkms\ for \thco (2--1)} { c} As a for \ceop. {
        Lowest contour level and increment 0.2\,\kkms\ } and { d}
      { As a for} CS (2--1) (in white) and CS (3--2) in red. {
        Lowest contours and increments are 0.2\,\kkms\ for CS (2--1)
        and 0.1\,\kkms\ for CS (3--2)}. }
 \label{figure:contours}
 \end{figure*}

The CO and CS spectra observed in the off-set map-position
+17\arcsec,10\arcsec\ are shown in Fig. \ref{fig:sample_spectra}. Only
components C2 and C3 are shown. Component C3 \twco and \thco (1--0)
line peak velocities are blueshifted with respect to the line peak
velocities of the other CO isotopologue and CS lines. This and the
slight \twco line asymmetry hints at self absorption. The redshifted
line wing suggests a weak outflow, but the line shape may also be due
to the CO self absorption.  Clear emission from the C2 component is
seen only in \twco and \thcop \ lines. The component is broad, flat
topped and the blueshifted \twco line wing hints at outflowing
material.

 The contours of component C3 integrated line emission of the three
 observed CO isotopologues  and the two CS transitions are shown in
 Fig. \ref{figure:contours} overlaid on the false colour VVV,
 \Halpha\ and Wise 12\,$\mum$ images. The CO (1--0) and CS (2-1)
 contours are white and the CO (2--1) and CS (3--2) contours red.  The
 \twco and \thco contours cover the whole globule. The integrated line
 emission in the \twco\ and \thco\ (1--0) transitions is distributed
 roughly spherically, but in the (2--1) transitions the maximum is
 closer to BHR\,160\,IRS1 and the core.  Both \ceo and CS transitions peak
 near the core. This can be explained by the higher critical densities
 of these molecules compared to \twcop. The highest transitions of all
 the observed molecules have higher critical depth than the lower
 transitions, and therefore they tend to peak more in the direction of
 the dense and obscured core.  \ceo and CS molecules are not excited
 in the low density envelope outside the core. Self absorption in
 the \twco line may also cause suppression of the \twco line in the
 core.

\begin{figure}
 \centering
  \includegraphics [width=8.8cm]{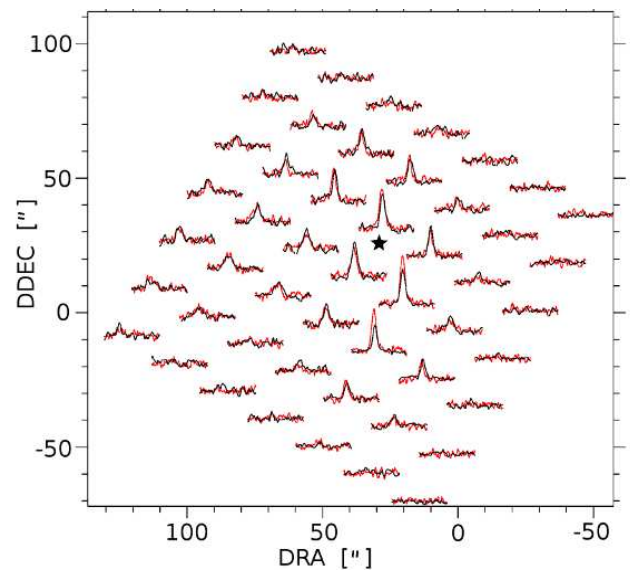}
    \caption{\ceo (1--0) and (2--1) (in red) spectra observed in
      BHR\,160. The spectra are in the \tastar scale and the maximum
      observed \ceo (2--1) intensity is 1.4K. The { asterisk indicates the  position of  BHR\,160\,IRS1.}}
 \label{figure:ceo}
 \end{figure}

The observed \ceo (1--0) and (2--1) spectra (in black and red,
respectively) in the \tastar scale are shown in
Fig. \ref{figure:ceo}. The \ceo (2--1) emission is stronger in two
positions in the core to the South of the map 0,0 position.  As the
SEST main beam efficiencies at the \ceo (1--0) and (2--1) frequencies
are 0.70 and 0.50, respectively, the difference in the intensity at
these positions { would be } even higher if expressed in the
\Tmb\ scale.
{ { However, the \Tmb\ scale is not useful when
  observing an extended object which has small scale structure that is
  smaller than the main beam.  The spatial resolution, especially in
  the \ceo (1--0) transition (Figs.  \ref{figure:contours} and
  \ref{figure:ceo}) is not sufficient to fully resolve the small scale
  structure in the BHR\,160 dense core but at the same time the lower
  intensity emission is more extended than the main beam. We note that
  the beams in Fig. \ref{figure:contours} indicate HPBW and not the
  main beam which is larger. Emission is therefore observed not only
  from the main beam but also from the beam side lobes and the error
  beam.  The SEST beam in the \ceo (2--1) transition traces better the
  dense core whereas the beam in the (1--0) transition, with an area
  four times that in the (2--1), also traces more of the less dense
  gas in the globule.  Direct comparison of the \ceo (1--0) and (2--1)
  brightness temperatures is meaningful only if the two transitions
  trace the same material.}} {  A likely explanation for the observed
  higher \tastar intensity of the \ceo (2--1) line with respect to the
  \ceo (1--0) line in the centre of the core is stronger beam dilution
  of the lower transition rather than a large difference in the
  excitation temperature. }

The main trend in the distribution of the observed molecules in
BHR\,160 is that the higher the  critical density is, the closer
the observed distribution is to the elongated core. Because of the
possible line self absorption, the distribution of the \twco emission
may be misleading.

\begin{figure*}
 \centering
 \includegraphics [width=18cm]{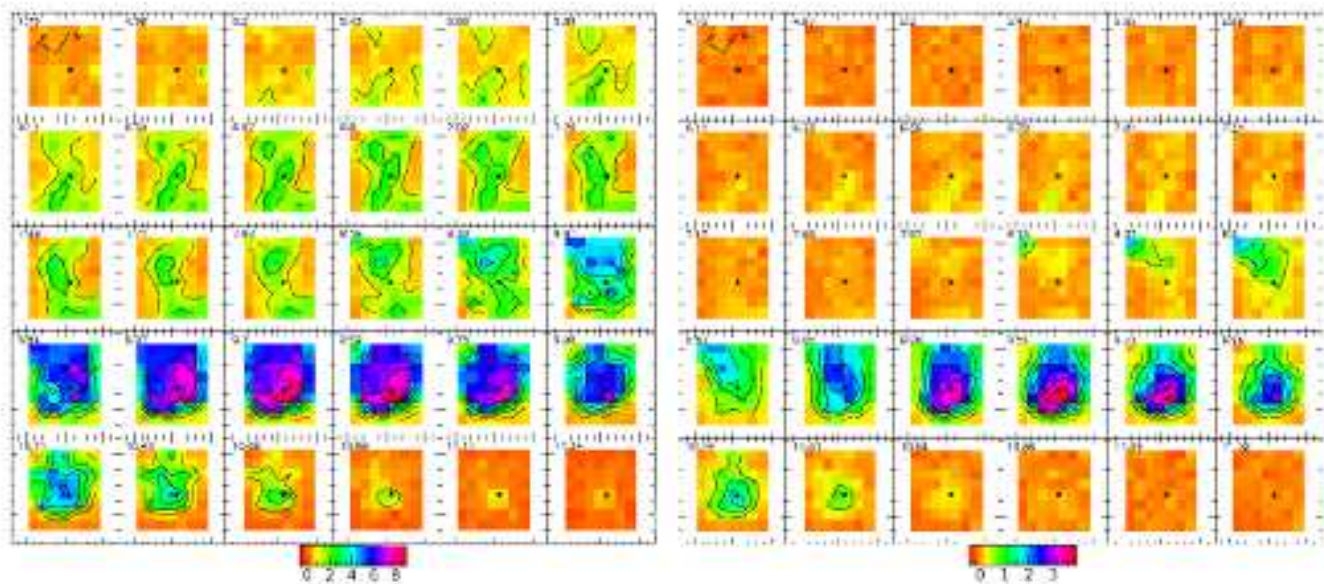}
    \caption{Channel maps of CO (2--1) (on the left) and \thco (2--1)
      (on the right) line emission in velocity bins of
      0.23~\kmps. Pixel scale is 20\arcsec\ by 20\arcsec \ and the
      North-East orientation is indicated in the upper left panel. The
      position of BHR\,160\,IRS1 is indicated by an asterisk in each
      panel.  { Lowest contour level and increment is 1.0\,\kkms\ 
      for \twcop\ and 0.5\,\kkms\  for \thcop. } }
\label{fig:CO-13CO-2-1-channel}
 \end{figure*}

The channel maps give more detailed information than the line
integrals. The \twco and \thco (2--1) and \twco (1--0), \thco (1--0),
\ceo (2--1), \thco (1--0), CS (3--2) and CS (2--1) channel maps are
shown in Fig. \ref{fig:CO-13CO-2-1-channel} and
Figs. \ref{fig:CO1-0_channel} to \ref{fig:CS3-2_channel},
respectively.

  As expected from the line integrals, the strongest
{  \twcop, \thcop, and \ceo (1--0) emission is seen in the direction of
  the globule centre and the IR source, whereas in the higher observed
  transitions the maximum emission is shifted towards the globule
  core.}

{
  The \ceo spectra (Fig.  \ref {figure:ceo}) and the \ceo (2--1) channel
  map (Fig. \ref{fig:C18O2-1_channel}), indicate that the
  maximum intensity which is seen at velocities 9.3 \kmps\ to 9.75
  \kmps, lies in the direction of the elongated core between the
  bright rim and the IR source. The signal-to-noise of the CS spectra
  is low, but the observed trend is similar to that observed in \ceo
  (2--1). The maximum emission in the integrated emission of CS (2--1)
  and (3--2) lines peaks in the core (Fig. \ref{figure:contours}\,C).
  It is reasonable to assume that the maximum density lies in the
  direction where the maximum \ceo (2--1) and CS line emission lies,
  that is in the direction of the obscured core. The spatial resolution
  of the present data is not sufficient to define the exact position
  and extent of the high density region. The maximum is only resolved
  in \ceo (2--1) line in the North-South direction but not in the
  direction perpendicular to it. The dimension of the core in the
  North-South direction is $\sim$1\arcmin. }
   
{ The \twco and \thco channel maps reveal yet another emission line
  component which emits at velocities between C2 and C3. The emission
  from this new component is seen in the Eastern corner of the mapped
  area. This line component is not evident in the spectrum  in Fig. \ref
  {fig:co2-1line} which is an average of \twco (2--1) lines around the
  centre of the map.}

\section{Discussion} \label{sect:discussion}

\subsection{Extinction} \label{sect:extinction}

The optical extinction, A$_V$, in the globule and its surroundings can be
estimated from the VVV DR2 and 2MASS photometry. { A method which
  is intermediate between the NICE \citep{ladalada1994} and NICER
  \citep{nicer} methods was used (see appendix \ref{sect:ext} for
details).}  The extinction, smoothed with a 40\arcsec\ Gaussian
  overlayed on the AAO false colour image, is shown in 
  Fig. \ref{fig:AAO_ext}.  The lowest contour is 5\minute\ and the
  increment is 1\minute. The extinction agrees well with the
  structures seen in the optical image.  Two compact areas are
  detected corresponding to BHR\,160 and { BHR\,159 }10\arcmin\ to
  the South of it.

\begin{figure}
  \includegraphics[width=8.8cm] {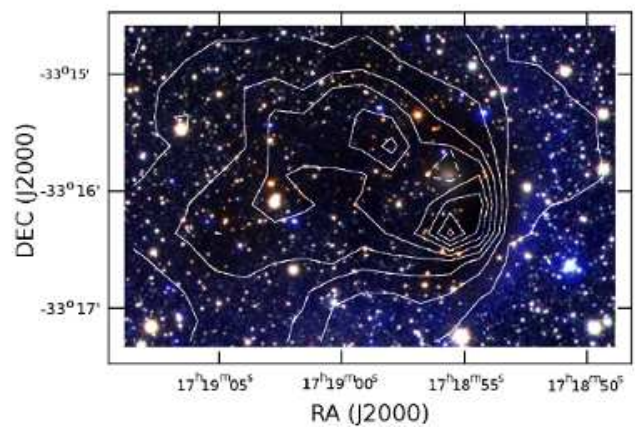}
    \caption{As Fig. \ref{figure:contours} with optical extinction 
      contours overplotted.  The contours start at 5\fm0  and
      the increment is 1\minute.}
\label{fig:BR1_ext}
 \end{figure}

Extinction contours, smoothed with a 30\arcsec\ Gaussian, in the
direction of BHR\,160 and overlayed on the VVV, \Halpha\ and WISE
false colour image are shown in Fig. \ref {fig:BR1_ext}. The lowest
contour is 5\minute\ and the increment is 1\minute. { The number of
  observed stars is low in the direction of the optically most opaque
  positions in the globule and if the map resolution is increased from
  30\arcsec, holes will appear into the map. The extinction minimum in
  the direction of BHR\,160\,IRS\,1 is likely due to high extinction at
  this position. This minimum is not seen in the large
  extinction map shown in Fig. \ref{fig:AAO_ext} where a
  40\arcsec\ smoothing Gaussian was applied.  The extinction agrees
  both with the obscuration seen in Fig. \ref{fig:BR1_ext} and with
  the observed distribution of molecular line emission
  (Fig. \ref{figure:contours})}.

The extinction reveals an elongated { {\it appendix} to the East of
  the globule main body. The extinction map is a two dimensional
  projection and traces the material along the line of sight. As BHR\,160
  is apparently associated with an extended compressed shell of a
  distant extinct HII region it is not unlikely that other structures
  in the shell are seen in projection in the same direction. Unlike
  extinction, the spectral line maps have an additional dimension,
  namely the velocity. The distribution of the emission from the C3 CO
  line component centred at 9.5\,\kmps\ is well correlated with the
  optical appearance of the main globule. As noted in Sect. \ref
  {sect:lines} there is indication of emission  at
  velocities between the C2 and C3 components in the \twco and \thco
  channel maps in the upper left corner of the panels. This
  corresponds to the very East corner of the mapped area adjacent to
  the {\it appendix}. The \thco (1--0) and (2--1) spectra in offset
  positions 0\arcsec,0\arcsec\ and +135\arcsec,10\arcsec\ are shown in
  the Fig. \ref{spectra-appendix}. The figure indicates that there is
  a separate velocity component centred at 8.5\,\kmps\ between
  components C2 and C3. The emission at 8.5\,\kmps\ coincides with the
  Western edge of the {\it appendix}.  The elongated cometary-like
  structures in the region (Figs. \ref{fig:ScoOB4} and \ref{fig:AAO})
  all face towards the O star HD 155806. However, different from this, the
  orientation of the {\it appendix} and BHR\,160 complex is elongated
  in the East-West direction. We argue thus that the {\it appendix} is
  not connected to BHR\,160 but is a separate structure seen in
  projection.}

\subsection{ The mass of BHR\,160} \label{sect:mass}

The cloud mass can be estimated from the observed \ceo line emission.
Assuming LTE, {an optically thin line and \ceo abundance of
  2.0$\times 10^{-7}$,} the maximum \Htwo\ column density {
  averaged over the SEST 23\arcsec\ beam } in the core is 4.4$\times
10^{22}$\,\persqcm\ if the \ceo excitation temperature is 10~K, and
3.7$\times 10^{22}$\,\persqcm\ if the temperature is 15-30~K. Assuming
LTE, an excitation temperature of 10\,K and an average molecular
weight of 2.8 per \Htwo\ molecule, the total integrated mass of the
globule summed up from the \ceo observations is {
  100~\Msun\ (d/1.6\,kpc)$^2$ where d is the distance to the
  globule. Approximately 70\% of the mass lies in the dense core. If
  the excitation temperature is higher, 15 to 30~K, the values are
  10\% lower. Besides the distance and excitation temperature also
  variations in the line optical depth and deviation from LTE cause
  further uncertainty to the mass estimation. A conservative estimate
  of the accuracy of the mass is $\pm$50\%. This mass refers also only
  to the molecular gas as traced by \ceo in the mapped area.} The
dense cloud core is surrounded by an extended less tenuous envelope
where \ceo is not excited.

 The optical extinction estimated from the VVV photometry offers an
 independent check on the \Htwo\ column density. Unlike molecular line
 or dust thermal emission, the extinction depends only on dust column
 density and not on gas/dust temperature nor density.  The maximum
 observed optical extinction for a single star in the core estimated
 from the VVV data is {20\minute$\pm$4\minute.} This value is a
 lower limit for the core maximum extinction as no stars are seen
 through the very centre of the core. Using the relation between the
 hydrogen column density and extinction in \citet{bohlinetal1978}, the
 estimated \Htwo\ column density { corresponding to this extinction
   is approximately $2.0 \pm 0.4 \times 10^{22}$\persqcm\ } (Av/E(B-V)
 is assumed to be 3.2). This is slightly lower than that derived in
 the centre of the core from the \ceo data. The mass of the globule
 { within the extinction contours of 7\minute\ in Fig.  \ref
   {fig:BR1_ext} is 210$\pm$80(d/1.6\,kpc)$^2$\,\Msun. The area is larger
   than that covered by the molecular line observations.  Divided
   between the {\it appendix} (area East of $17\hour 19\minute$\ and
   South of $-33\degr 15\arcmin 20\arcsec$) and the globule the masses
   are 80$\pm$30\,\Msun (d/1.6\,kpc)$^2$\ and 130$\pm$50\,\Msun
   (d/1.6\,kpc)$^2$, respectively.  Here it is assumed that the
   Gaussian smoothed extinction is accurate to 3.0\minute\ in the
   direction of the globule and 2.0\minute\ in the {\it appendix}
   (chiefly Fig. \ref{fig:BR1_ext_rms}). }

We have also done radiative transfer modelling using the Monte Carlo
method \citep{juvela1997} assuming a centrally condensed, spherically
symmetric geometry. Even though the core is clearly elongated, the
modelling should give an indication of the possible \Htwo\ densities
and the excitation temperature in the core.  Only the C3 line
component and { the least optically thin lines, \ceo (2--1)/(1--0)
  and lines tracing density (CS (3--2)/(2--1), were considered.) } The
excitation temperature in the centre of the core had to be raised to
40\,K to explain the observed \ceo (2--1)/(1--0) line ratio.  In the
outer parts, the temperature dropped to 20\,K. The model which could
reproduce the observed \ceo and CS lines had a central density of
3$\times 10^{4}$\,\percubcm\ which rapidly decreased and was less than
1000\,\percubcm\ at the edges.  The total mass of such a model is 100
\Msun.  {These values are in line with those derived from the LTE
  approximation. { However,} this should be only considered to
  demonstrate that it is possible to construct a hypothetical cloud
  core which reproduces the observed line intensities using similar
  input values as those derived from the LTE approximation.}

\subsection{Star formation} \label{sect:starformation}

\begin{figure}
 \centering
 \includegraphics [width=8.8cm]{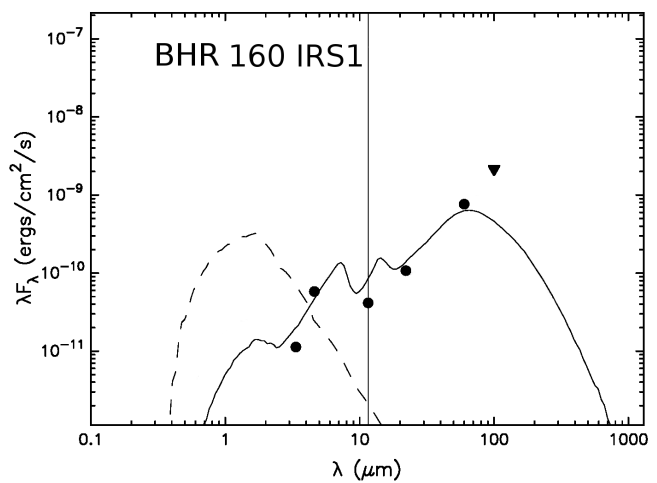}
 \caption{SED fit to BHR\,160 IRS1 WISE and IRAS data. The IRAS
   100\,$\mum$ value is an upper limit. The continuous line shows the
   fit and the dashed line the  SED of the {  unreddened} central source.}
 \label{figure:BR1SED}
 \end{figure}

The only IR excess source detected so far within BHR\,160 is IRS1
(WISE J171855.54-331554.3). We have used the spectral energy
distribution (SED) fitting tool of \citet{robitailleetal2007} to fit
the WISE MIR and IRAS FIR data (Fig. \ref{figure:BR1SED}). As the
IRAS\,100\,$\mum$ flux is likely to contain emission from the globule
dense core, it is given as an upper limit.  The data wavelength range
is not sufficient for a unique fit but the results point to a Class\,I
protostar.  The fitted stellar temperature is 4400\,K, with masses and
luminosity range between 1.5 and 2.5\,\Msun\ and 70 to 200\,\Lsun.
However, even a five \Msun\ high mass star is suggested by one fit.

SED fits were also made for the seven additional stellar sources marked in
Fig.\,\ref{fig:WISE}. No \jhks\ measurements are available for source
7 as it is not visible even in the VVV data so only WISE data are
used.  2MASS photometry was used for the remaining sources because the
stars are saturated in the VVV photometry.  As in the case of
BHR\,160\,IRS1, no unique fit is possible for most of the sources, but
the fits are sufficient to demonstrate that the objects are likely to be 
young stars with a dust shell and/or disk. Even though the fitted
values should be considered only as indicative, the result is
sufficient to show that active star formation is taking place in the
region.  The results of the fits are listed in
Table\,\ref{TABLE:sed_fits} and plotted in the  
Fig.\,\ref{fig:sed1}.  It would be possible to fit most of the sources
assuming strong interstellar reddening only, but in that case the
distance to the stars would have to be less than 20~pc. The fitted Av
of 15\umag\ to 30\umag\ is not plausible for such nearby sources.

BHR\,160\,IRS1 is associated with a dense circumstellar envelope. It
is, however, not clear how much the BHR\,160\,IRS1 IRAS fluxes are
contaminated by the adjacent globule core.  Except for star 7, the
remaining stars in Table\,\ref{TABLE:sed_fits} are hotter, much more
luminous and have higher mass (M $>$\ 4\Msun) than BHR\,160\,IRS1.
The projected distance on the sky between BHR\,160\,IRS1 and the
fitted stars is between 0.5 and 2.4\,pc. It is likely that the stars
are also newly formed out of the remnant neutral material surrounding
the Sco OB4 association.

\begin{table}
  \caption[]{SED fits}
   \label{TABLE:sed_fits}
\begin{tabular}{ccccc}
\hline
\hline
star &    WISE  &         T      &    M    & L       \\
     &          &       ($\times10^3$K)& (\Msun) & (\Lsun)  \\
\hline
1 & J171855.54-331554.3 & 4.4   &  1.8-5.6  & 70-200    \\
2 & J171917.36-331941.3 & 15-20 &  4.8-6.5  & 930-1500  \\ 
3 & J171913.86-331917.7 & 18-26 &  5.0-10.0 & 1000-2000 \\
4 & J171900.44-331734.9 &  25   &  9.6      & 5500      \\
5 & J171855.17-331716.0 &  15   &  4.5-5.5  & 2700-4000 \\
6 & J171854.59-331747.9 &  15   &  4.4      & 2500      \\
7 & J171833.24-331538.6 &  4    &  0.7-2    & 10-30   \\
8 & J171833.26-331538.5 &  25   &  9.7      & 5500      \\
\hline
\end{tabular}                                          
\end{table}

\subsection{Bright rim}
The intense bright rim, catalogued as \object {Magakian 714}
\citep{magakian2003}, that is associated with BHR\,160 is very rare
among cometary globules. The only case that we are aware of with a
similar intense bright rim is the globule \object {CG\,1} in the Gum Nebula,
which is illuminated by the young Herbig Ae/Be star NX~Pup =
Cod--44$^\circ$3318, born within the globule \citep{reipurth1983,
  haikalaetal2010}. We here consider the relationship between BHR\,160
and HD~319648, which is located just 30" in front of the globule.

Noel Cramer of Geneva Observatory has kindly obtained Geneva
photometry of HD~319648, and using the calibrations of \citet{cramer1993} he
has derived the following parameters: m$_V$=10.593, M$_V$=-1.7,
A$_V$=1.18, distance 1.64~kpc, and a spectral type of B3III. The
distance is in excellent agreement with the distance of 1.61~kpc
derived by \citet{persitapia2008} for the nearby NGC~6334 cluster,
which is assumed to be also part of the Sco~OB4 association
\citep{roslund1966}. The spectral type indicates that the star has enough UV
radiation to form a photoionized region in the Magakian~714 bright
rim. The projected separation of HD~319648 and the bright rim is only
0.23~pc. We thus consider it highly likely that the globule and the star are
located close to each other, and that their closeness in the sky is
not just a line-of-sight effect. The question is then whether
HD~319648 was born in the globule or just is passing by the globule.
Cases of random encounters between stars and globules are known, a
fine case is the globule CG13 in the Gum Nebula, which has encountered
a star passing through it \citep{reipurth1983}.

HD~319648 is clearly a member of the Sco OB4 association \citep{roslund1966}
but could have been formed elsewhere in the association and by
chance moved into the vicinity of BHR\,160. The mean velocity
dispersion of OB stars in Sco~OB4 is $\sim$7~km~s$^{-1}$ \citep{roslund1969},
and a lower limit to the time HD~319648 could have been in the
vicinity of the globule comes from dividing the length of the bright
rim ($\sim$2$'$) with this velocity, which yields a time span of
130,000~yr. Since the Sco OB4 association must be at least several
million years old, it is certainly possible that one of the
association members during that time has wandered into the vicinity of
BHR\,160.  However, the fact that the star is placed right in front of
the globule suggests to us that it
may have formed in the globule as a secondary star formation event,
probably triggered by the massive O star HD~155806, which is also
located along the main axis of the globule. We may therefore be
witnessing a case of sequential star formation, where the formation of
HD 319648 was triggered by HD~155806, while HD~319648 in turn may have
triggered the formation of the currently embedded source IRS~1 within the
globule.

\subsection{Cloud structure} \label{sect:structure}

 Biologically BHR\,160 is a scaled up version of the head of the
 archetype cometary globule CG\,1
 \citep{HawardenBrand1976,reipurth1983} in the Gum Nebula.  The CG\,1
 bright rimmed head has a diameter of a few arcminutes and is
 illuminated by the nearby pre-main-sequence star NX Pup, and a
 protostar, CG\,1\,IRS\,1, is situated at the side of the elongated
 core \citep{haikalaetal2010}.  The total \ceo mass of the CG\,1 core
 is, however, only a tenth of that measured in  BHR\,160. And another
 difference is that CG\,1\,IRS1 is situated between the illuminating
 star and the core whereas in BHR\,160 the protostar is on the
 opposite side of the core.
 The similarities in the morphology and the location of these two
 globules at the edge of a large, well evolved HII region suggest a
 similar origin. In \citet{makelahaikala2013} it was argued that the
 formation of CG\,1 was driven by radiation induced implosion \citep
 {reipurth1983}. Less dense material extends to NE from the dense
 BHR\,160 core, but there is no indication in Fig.\ \ref {fig:AAO_ext}
 that the globule has a long tail like that of CG\,1. The cometary
 globule (known as BHR\,159) just south of BHR\,160 also points at similar
 processes being active here and in the Gum Nebula.

\section{Summary and Conclusions} \label{sect:conclusions}

We have carried out dedicated molecular line observations in the CO
(and isotopologues) (1--0), (2--1) and in CS (2--1) and (3--2)
transitions to study the basic properties of the bright rimmed dark
cloud BHR\,160. Combining these data with data available in various
public surveys provides insight into the globule and its surroundings:

1.  AAO/UKST imaging in the \Halpha\ line reveals that BHR\,160 is part
of the shell of an extinct HII region. Besides BHR\,160 several other
globules are associated with the same shell of neutral material at the
edge of the Sco OB4 association.

2. BHR\,160 is elongated with dimensions of approximately
5\arcmin$\times$3\arcmin. The globule is delineated on three sides by
a halo emitting in the \Halpha\ line while a cometary-like tail is
present on the NE side.

3.  The 1\arcmin\ by 0\farcm6 BHR\,160 core is dense and bound sharply in
the West by the bright \Halpha\ rim which is  also seen faintly in reflected
light at other wavelengths.

4. The maximum \Htwo\ column density in the globule core is
$4.4\times10^{22}$\,\persqcm\ as estimated from the \ceo data, or half
of this when estimated from extinction. Mass estimated from the \ceo
observations { using LTE approximation within the mapped region is
  100$\pm$50(d/1.6\,kpc)$^2$\,\Msun\ of which 70\% is contained in the
  core. The total BHR\,160 mass estimated from optical extinction is
  210$\pm$80(d/1.6\,kpc)$^2$\,\Msunp\ Approximately 40\% of the total
  optical mass is contained in the BHR\,160 appendix which is argued
  to be a separate cloud seen in projection.}

5. Only one infrared excess star, BHR\,160\,IRS1, was detected within
BHR\,160.  It lies off the core on the opposite side to the bright
rim. Only a faint slightly extended object is seen in the VVV
\Ks\ image but the star is seen in all the four WISE channels from
3.4~$\mum$ to 20~$\mum$.  An IRAS source lies nominally 15\arcsec\ to
the West of IRS\,1, but in the HIRES enhanced IRAS 60\,$\mum$\ image the
emission maximum coincides with IRS\,1. A SED fit to the WISE
and IRAS 60\,$\mum$\ fluxes points at a low mass Class I protostar of
about 2~\Msun. Other fits from subsolar up to  5~\Msun\ are, however,
possible and the fits must be considered only indicative until better
FIR data become available.

5. Analysis of seven bright stellar WISE sources seen towards BHR\,160
provides evidence for recent star-formation in the region. According
to SED fits to \jhks\ and WISE data, six of these are high mass
objects in different stages of early stellar evolution. Only one low
mass source is detected, but only the WISE fluxes are available for
this star. As all the fits are based on a restricted wavelength range
they should all be considered as indicative.

\vspace{0.5cm}

{\em Acknowledgements:} We thank Noel Cramer for obtaining the Geneva
photometry of HD~319648, and Bambang Hidayat for sending us a copy of
the Th\'e~(1961) paper.  This research has made use of the following
resources: data products from observations made with ESO Telescopes at
the La Silla or Paranal Observatories under ESO programme ID
179.B-2002; SIMBAD database, operated at CDS, Strasbourg, France;
NASA’s Astrophysics Data System Bibliographic Services; data from the
Southern  H$\alpha$\  Sky Survey Atlas (SHASSA), which was produced with support
from the National Science Foundation; data products from the 2MASS,
which is a joint project of the University of Massachusetts and the
Infrared Processing and Analysis Center/California Institute of
Technology, funded by the NASA and the US National Science Foundation;
data products from the Wide-field Infrared Survey Explorer, which is a
joint project of the University of California, Los Angeles, and the
Jet Propulsion Laboratory/California Institute of Technology, funded
by the National Aeronautics and Space Administration.

%

\bibliographystyle{aa}
\bibliography{AA201526777.bib}

\begin{thebibliography}{28}
\expandafter\ifx\csname natexlab\endcsname\relax\def\natexlab#1{#1}\fi

\bibitem[{Auman {et~al.}(1990)Auman, Fowler, \& Melnyk}]{aumanetal1990}
Auman, H., Fowler, J., \& Melnyk, M. 1990, AJ, 99, 1674

\bibitem[{{Bohlin} {et~al.}(1978){Bohlin}, {Savage}, \&
  {Drake}}]{bohlinetal1978}
{Bohlin}, R.~C., {Savage}, B.~D., \& {Drake}, J.~F. 1978, \apj, 224, 132

\bibitem[{{Bok}(1977)}]{Bok1977}
{Bok}, B.~J. 1977, \pasp, 89, 597

\bibitem[{{Bourke} {et~al.}(1995{\natexlab{a}}){Bourke}, {Hyland}, \&
  {Robinson}}]{bourkeetal1995a}
{Bourke}, T.~L., {Hyland}, A.~R., \& {Robinson}, G. 1995{\natexlab{a}}, \mnras,
  276, 1052

\bibitem[{{Bourke} {et~al.}(1995{\natexlab{b}}){Bourke}, {Hyland}, {Robinson},
  {James}, \& {Wright}}]{bourkeetal1995b}
{Bourke}, T.~L., {Hyland}, A.~R., {Robinson}, G., {James}, S.~D., \& {Wright},
  C.~M. 1995{\natexlab{b}}, \mnras, 276, 1067

\bibitem[{{Cramer}(1993)}]{cramer1993}
{Cramer}, N. 1993, \aap, 269, 457

\bibitem[{{Foster} {et~al.}(2008){Foster}, {Rom{\'a}n-Z{\'u}{\~n}iga},
  {Goodman}, {Lada}, \& {Alves}}]{fosteretal2008}
{Foster}, J.~B., {Rom{\'a}n-Z{\'u}{\~n}iga}, C.~G., {Goodman}, A.~A., {Lada},
  E.~A., \& {Alves}, J. 2008, \apj, 674, 831

\bibitem[{{Fullerton} {et~al.}(2011){Fullerton}, {Petit}, {Bagnulo}, {Wade}, \&
  {Wade}}]{fullertonetal2011}
{Fullerton}, A.~W., {Petit}, V., {Bagnulo}, S., {Wade}, G.~A., \& {Wade}. 2011,
  in IAU Symposium, Vol. 272, IAU Symposium, ed. C.~{Neiner}, G.~{Wade},
  G.~{Meynet}, \& G.~{Peters}, 182--183

\bibitem[{{Haikala} {et~al.}(2010){Haikala}, {M{\"a}kel{\"a}}, \&
  {V{\"a}is{\"a}nen}}]{haikalaetal2010}
{Haikala}, L.~K., {M{\"a}kel{\"a}}, M.~M., \& {V{\"a}is{\"a}nen}, P. 2010,
  \aap, 522, A106

\bibitem[{{Haikala} \& {Reipurth}(2010)}]{haikalareipurth2010}
{Haikala}, L.~K. \& {Reipurth}, B. 2010, \aap, 510, A1

\bibitem[{{Hartley} {et~al.}(1986){Hartley}, {Tritton}, {Manchester}, {Smith},
  \& {Goss}}]{hartleyetal1986}
{Hartley}, M., {Tritton}, S.~B., {Manchester}, R.~N., {Smith}, R.~M., \&
  {Goss}, W.~M. 1986, \aaps, 63, 27

\bibitem[{Hawarden \& Brand(1976)}]{HawardenBrand1976}
Hawarden, T. \& Brand, P. 1976, MNRAS, 175, 19P

\bibitem[{{Juvela}(1997)}]{juvela1997}
{Juvela}, M. 1997, \aap, 322, 943

\bibitem[{Lada {et~al.}(1994)Lada, Lada, Clemens, \& Bally}]{ladalada1994}
Lada, C., Lada, E., Clemens, D., \& Bally, J. 1994, \apj, 429, 694

\bibitem[{Lombardi \& Alves(2001)}]{nicer}
Lombardi, M. \& Alves, J. 2001, A\&A, 377, 1023

\bibitem[{{Magakian}(2003)}]{magakian2003}
{Magakian}, T.~Y. 2003, \aap, 399, 141

\bibitem[{{M{\"a}kel{\"a}} \& {Haikala}(2013)}]{makelahaikala2013}
{M{\"a}kel{\"a}}, M.~M. \& {Haikala}, L.~K. 2013, \aap, 550, A83

\bibitem[{{Parker} {et~al.}(2005){Parker}, {Phillipps}, {Pierce}, {Hartley},
  {Hambly}, {Read}, {MacGillivray}, {Tritton}, {Cass}, {Cannon}, {Cohen},
  {Drew}, {Frew}, {Hopewell}, {Mader}, {Malin}, {Masheder}, {Morgan}, {Morris},
  {Russeil}, {Russell}, \& {Walker}}]{parkeretal2005}
{Parker}, Q.~A., {Phillipps}, S., {Pierce}, M.~J., {et~al.} 2005, \mnras, 362,
  689

\bibitem[{{Persi} \& {Tapia}(2008)}]{persitapia2008}
{Persi}, P. \& {Tapia}, M. 2008, {in Handbook of Star Forming Regions, Volume
  II: Astron. Soc. Pacific, ed. Bo Reipurth}, 456

\bibitem[{{Reipurth}(1983)}]{reipurth1983}
{Reipurth}, B. 1983, \aap, 117, 183

\bibitem[{{Reipurth}(2008)}]{reipurth2008}
{Reipurth}, B. 2008, {in Handbook of Star Forming Regions, Volume II: Astron.
  Soc. Pacific, ed. Bo Reipurth}, 847

\bibitem[{{Robitaille} {et~al.}(2007){Robitaille}, {Whitney}, {Indebetouw}, \&
  {Wood}}]{robitailleetal2007}
{Robitaille}, T.~P., {Whitney}, B.~A., {Indebetouw}, R., \& {Wood}, K. 2007,
  \apjs, 169, 328

\bibitem[{{Roslund}(1966)}]{roslund1966}
{Roslund}, C. 1966, Arkiv f\"or Astronomi, 4, 101

\bibitem[{{Roslund}(1969)}]{roslund1969}
{Roslund}, C. 1969, Arkiv f\"or Astronomi, 5, 209

\bibitem[{{Saito} {et~al.}(2012){Saito}, {Hempel}, {Minniti}, {Lucas},
  {Rejkuba}, {Toledo}, {Gonzalez}, {Alonso-Garc{\'{\i}}a}, {Irwin},
  {Gonzalez-Solares}, {Hodgkin}, {Lewis}, {Cross}, {Ivanov}, {Kerins},
  {Emerson}, {Soto}, {Am{\^o}res}, {Gurovich}, {D{\'e}k{\'a}ny}, {Angeloni},
  {Beamin}, {Catelan}, {Padilla}, {Zoccali}, {Pietrukowicz}, {Moni Bidin},
  {Mauro}, {Geisler}, {Folkes}, {Sale}, {Borissova}, {Kurtev}, {Ahumada},
  {Alonso}, {Adamson}, {Arias}, {Bandyopadhyay}, {Barb{\'a}}, {Barbuy},
  {Baume}, {Bedin}, {Bellini}, {Benjamin}, {Bica}, {Bonatto}, {Bronfman},
  {Carraro}, {Chen{\`e}}, {Clari{\'a}}, {Clarke}, {Contreras}, {Corvill{\'o}n},
  {de Grijs}, {Dias}, {Drew}, {Fari{\~n}a}, {Feinstein},
  {Fern{\'a}ndez-Laj{\'u}s}, {Gamen}, {Gieren}, {Goldman},
  {Gonz{\'a}lez-Fern{\'a}ndez}, {Grand}, {Gunthardt}, {Hambly}, {Hanson},
  {He{\l}miniak}, {Hoare}, {Huckvale}, {Jord{\'a}n}, {Kinemuchi}, {Longmore},
  {L{\'o}pez-Corredoira}, {Maccarone}, {Majaess}, {Mart{\'{\i}}n}, {Masetti},
  {Mennickent}, {Mirabel}, {Monaco}, {Morelli}, {Motta}, {Palma}, {Parisi},
  {Parker}, {Pe{\~n}aloza}, {Pietrzy{\'n}ski}, {Pignata}, {Popescu}, {Read},
  {Rojas}, {Roman-Lopes}, {Ruiz}, {Saviane}, {Schreiber}, {Schr{\"o}der},
  {Sharma}, {Smith}, {Sodr{\'e}}, {Stead}, {Stephens}, {Tamura}, {Tappert},
  {Thompson}, {Valenti}, {Vanzi}, {Walton}, {Weidmann}, \&
  {Zijlstra}}]{saitoetal2012}
{Saito}, R.~K., {Hempel}, M., {Minniti}, D., {et~al.} 2012, \aap, 537, A107

\bibitem[{{Th\'e}(1961)}]{the1961}
{Th\'e}, P.-S. 1961, Contributions from the Bosscha Observervatory, 12, 0

\bibitem[{{Walborn}(1973)}]{walborn1973}
{Walborn}, N.~R. 1973, \aj, 78, 1067

\bibitem[{{Wright} {et~al.}(2010){Wright}, {Eisenhardt}, {Mainzer}, {Ressler},
  {Cutri}, {Jarrett}, {Kirkpatrick}, {Padgett}, {McMillan}, {Skrutskie},
  {Stanford}, {Cohen}, {Walker}, {Mather}, {Leisawitz}, {Gautier}, {McLean},
  {Benford}, {Lonsdale}, {Blain}, {Mendez}, {Irace}, {Duval}, {Liu}, {Royer},
  {Heinrichsen}, {Howard}, {Shannon}, {Kendall}, {Walsh}, {Larsen}, {Cardon},
  {Schick}, {Schwalm}, {Abid}, {Fabinsky}, {Naes}, \& {Tsai}}]{wrightetal2010}
{Wright}, E.~L., {Eisenhardt}, P.~R.~M., {Mainzer}, A.~K., {et~al.} 2010, \aj,
  140, 1868

\end{thebibliography}

\begin{appendix}

{
  \section{Estimation of the extinction: SIMPLE} \label{sect:ext}

{ The spread of the (\J--\H) and (\H--K) colour indices of unreddened main
 main sequence stars is small, being approximately 0\fm7 and 0\fm5,
 respectively. For the giant branch the spread of the (\H--\K) index is even
 smaller (0\fm3). Interstellar reddening increases both colour
 indices by an amount which is proportional to the amount of the
 reddening. If the reddening law is known it is possible to estimate
 the reddening affecting each measured star from its position relative to
 the unreddened main or giant sequence.
 A similar method was used traditionally for example when estimating the
 interstellar reddening of stars using the \UBV colour-colour diagram.
 When the NIR photometry was conducted using a single pixel detector it was
 possible to estimate the reddening of each observed star
 manually. However, the number of stars detected using modern NIR
 arrays is so large that estimating the reddening manually star by
 star is not feasible and a more automated method must be
 applied.

 \citet{ladalada1994} introduced the NICE (Near Infra Red Excess)
 method to automatically estimate the the reddenings from stellar NIR
 colours using the observed (\H--\K) colour index.  Intrinsic
 (\H--\K)$_o$\ values are  estimated from the colour distribution of the
 stellar population in a non-reddened control field near the programme
 field. Any deviation from the intrinsic colour is attributed to
 reddening. NICER (Near Infra Red Excess Revisited) method
 \citep{nicer} utilises both the (\J--\H) and (\H--\K) colour
 indices. In addition to intrinsic (\H--\K)$_o$\ and
 (\J--\H)$_o$\ indices also the colour distribution of the stellar
 population is estimated using a non reddened control field. The NICE
 and NICER provide the difference in reddening between the programme and
 control fields.  For a summary of the two methods see \cite{nicer}.

NICE or NICER methods work best when observing nearby dark clouds off
the Galactic plane where the interstellar reddening outside the cloud
is small.  Both methods are problematic if the programme
field lies in a direction which contains many dust clouds, possibly
both in front and behind the object, and no nearby unreddened control
field can be found.  This is typically the case especially in or near
the Galactic plane in the general direction of the Galactic centre.

BHR\,160 lies near the Galactic plane only $\sim$7\degr\ from the
centre ({\it l=}353.2518\degr, {\it b=}+02.4154\degr) at a distance of
1.6\,kpc.  Any line of sight in this direction is bound to contain
unrelated molecular clouds and diffuse ISM.  No nearby unreddened
control field could be found.  Therefore a robust method, which is 
basically automatising the traditional method of estimating reddenings
from the \jhks\ colour-colour diagram was applied. This method, which we call
SIMPLE, is less sophisticated than the NICER method but is sufficient
for the purpose of this paper.

Contrary to NICER the zero point for the reddening is assumed, not
estimated from the distribution of stars in a control field.  A
(\J--\H)$_o$--(\H--\K)$_o$\ {  value 0\fm8, 0\fm16 below the unreddened late main
sequence which is typical for an unreddened Galactic stellar field
was adopted.}  The actual field stellar
population depends on the Galactic line of sight (Galactic pole,
anti-centre, centre or intermediate) which leads to an additional
error source as compared with the NICER method. The reddening for each
star is estimated as in the NICER method and corresponds to the
distance from the star along the reddening line to a line
perpendicular to this, the zero line, which goes through the adopted
(\J--\H)$_o$--(\H--\K)$_o$\ position.  It is impossible to decide only
from stellar \jhks \ colours if a star is a late-type main sequence
star or an early-type star.  The stars below the zero line are,
however, likely to be early type stars. Instead of zeroing their
reddening they are assumed to originate from the earlier main sequence
and their reddening is estimated. The reddening for stars near the
unreddened early main sequence was set to zero. The imaged area is
divided into bins where the reddening is estimated as the Gaussian
weighted median of stars near the selected position. If only one or no
stars is found the pixel is blanked. The adopted pixel size is half
the Gaussian half width. { The estimate is the average extinction in the
pixel and can be much smaller than the maximum extinction detected in
it.}

 To estimate the interstellar reddening in the direction of BHR\ 160
using the VVV data base, objects classified as stars (VVV ‘merged
class 1’) detected in all \jhks\ bands and with a magnitude error less
than 0\fm2 were included. In addition VVV stars fainter than
17\minute\ in J band were excluded and stars brighter than 12\minute\
in J band were replaced by 2MASS photometry. Only stars within or near
the reddening line were used. This selects stars with intrinsic colours
similar to main sequence stars and giants and eliminates outliers
which consist mainly of stars with strong NIR excess and a variety of
evolved stars, galaxies and AGNs \citep [see
for example][]{fosteretal2008}. The number of objects in the area imaged in Fig
3 and full-filling these requirements is 51000. The
(\J--\H)--(\H--\K) colour-colour diagram of the selected stars is shown
in Fig. \ref {figure:vvvselection}. The stars in the three regions
described above are colour coded by green (zero reddening stars near
the main sequence), red (early type stars below the late main sequence) and
black (stars in or near the reddening line).

{
\begin{figure}[h]
  \centering
  \includegraphics [width=8.8cm]{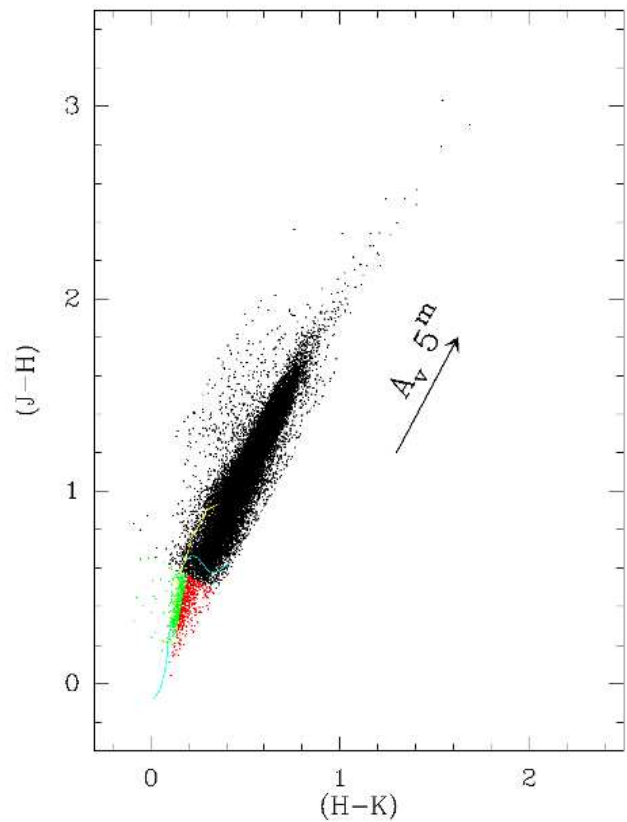}
  \caption{J--H, H--K colour colour diagram of the VVV stars in Fig. \ref {figure:BR1VVV} fullfilling the selection criteria.  The blue and yellow curves
    indicate the locations of unreddened dwarf and giant stars. The arrow indicates the effect of  5\umag\ of reddening. The stars assumed to be unreddened main sequence stars and the assumed reddened early type stars below the late main sequense are plotted in green and red, respectively }
   \label{figure:vvvselection}
\end{figure}
 }

 Almost any pixel in the area being studied contains stars with zero
 or near zero reddenings and also reddenings significantly larger than
 the average. The standard deviation of reddenings included into the
 average is thus large and applying sigma clip would mask both the
 near zero and the large reddenings. However, the large reddenings in
 the inspected area are most probably correct and trace the densest
 clumps in the area. Masking these stars would therefore result in
 underestimating reddening in the clouds. Therefore instead of
 applying sigma-clip it was decided only to mask the reddenings
 clearly smaller than the median of the stars in each pixel. BHR\,160
 lies 1.6 kpc away and the near zero reddenings trace the nearby
 stars. Because of this choice the resulting rms (noise estimate) of
 reddenings is large in directions to the dense clumps in the region.
 The calculated noise estimates in the area of Fig. \ref {fig:AAO_ext}
 is shown in Fig. \ref {fig:AAO_noise}. The estimated noise is below
 two magnitudes over most of the image area and as expected, rises in
 the regions of high extinction. The locations of high extinction tend
 to stand out from the surroundings better in the noise images than in
 the extinction images.  { The noise is highest in the direction of
 BHR\,159 and BHR 160.}

  \begin{figure}[h]
  \centering
  \includegraphics [width=8.8cm]{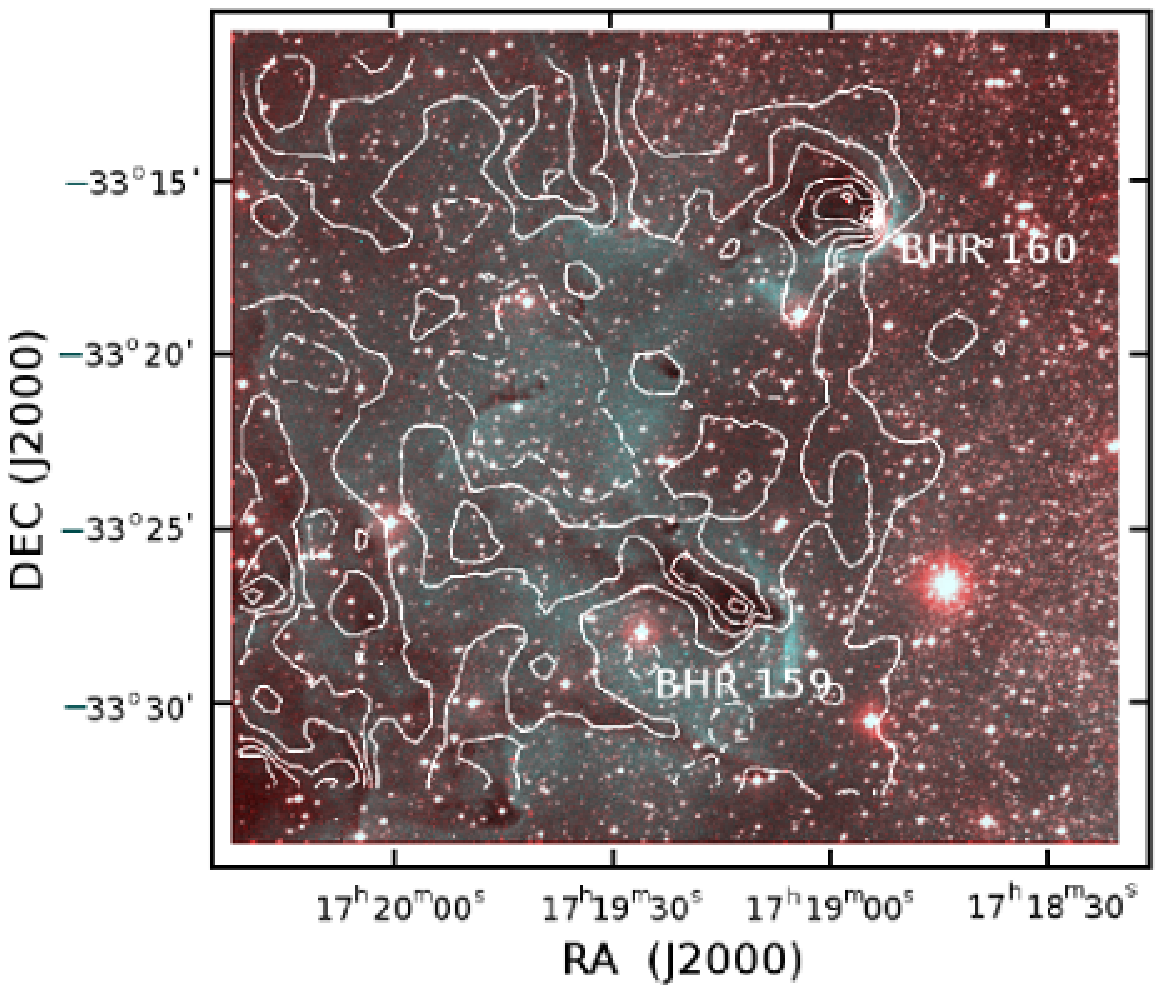}
   \caption{As Fig. \ref{fig:AAO} with extinction contours
      overlayed. The contours range from 5\umag \ to 12\umag 
     in steps of 1\fm0.}
   \label{fig:AAO_ext}
  \end{figure}

 { The accuracy of the derived reddening distributions depends mainly on
 the limiting magnitude of the stellar photometry data and stellar
 density in the programme field. The lower the limiting magnitude the
 lower is the maximum reddening that one can detect. The spatial
 resolution of the derived reddening distribution is proportional to
 the stellar density. The more stars are observed per square arcminute
 the smaller the selected pixel size can be. The accuracy of the
 resulting reddening distribution is a combination of these two
 effects. The maximum possible spatial resolution and the accuracy of
 the reddening distribution varies within the field. The noise
 estimate (Fig. \ref {fig:AAO_noise}) is low in areas of low
 extinction as many stars are detected per pixel (Fig. \ref
 {figure:vvvselection}). Because of the high extinction in the
 direction of BHR\,159 and BHR\,160 only few stars can be seen through
 the centres of the globules. The reddening is too high for the
 distant late main sequence stars to be detected (limiting magnitude
 of the photometric data) through the globules and thus the nose
 estimate is high. Only few highly reddened stars can be seen. When
 the extinction is too high no background stars can be seen. This is
 probably the case in the direction of BHR\,160\,IRS1 where a local
 minimum is seen in the reddening distribution (Fig. \ref
 {fig:BR1_ext}). The number of stars in the colour-colour diagram in
 Fig.\ref {figure:vvvselection} decreases significantly when the
 estimated extinction is nine magnitudes or more.  The areas in
 Figs. \ref {fig:BR1_ext} and \ref{fig:AAO_ext} where the extinction
 is higher than 9\umag\ are likely to be lower limits.  }

 }
 
  {
  \begin{figure} [h]
  \centering
  \includegraphics [width=8.8cm]{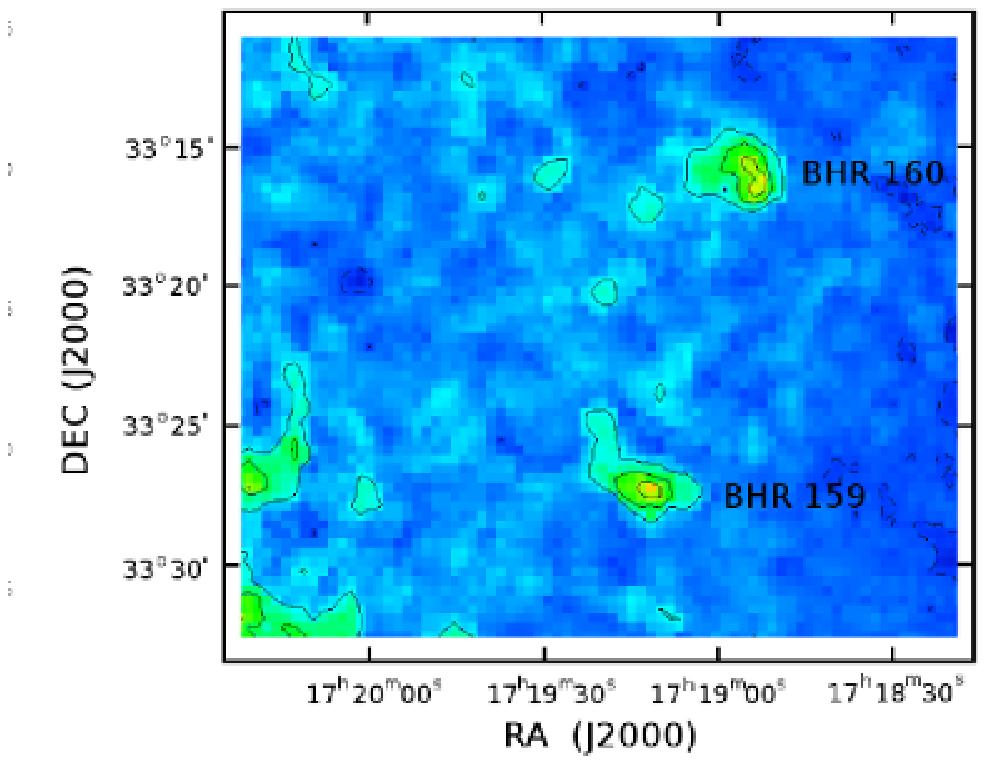}
  \caption{Noise estimate for the extinctions in Fig.
    \ref {fig:AAO_ext}. The contours range from 1\umag\ (dashed) to 4\umag 
      in steps of  1\fm0.}
   \label{fig:AAO_noise}
\end{figure}
 }

  {
\begin{figure}[h]
  \centering
  \includegraphics [width=8.8cm]{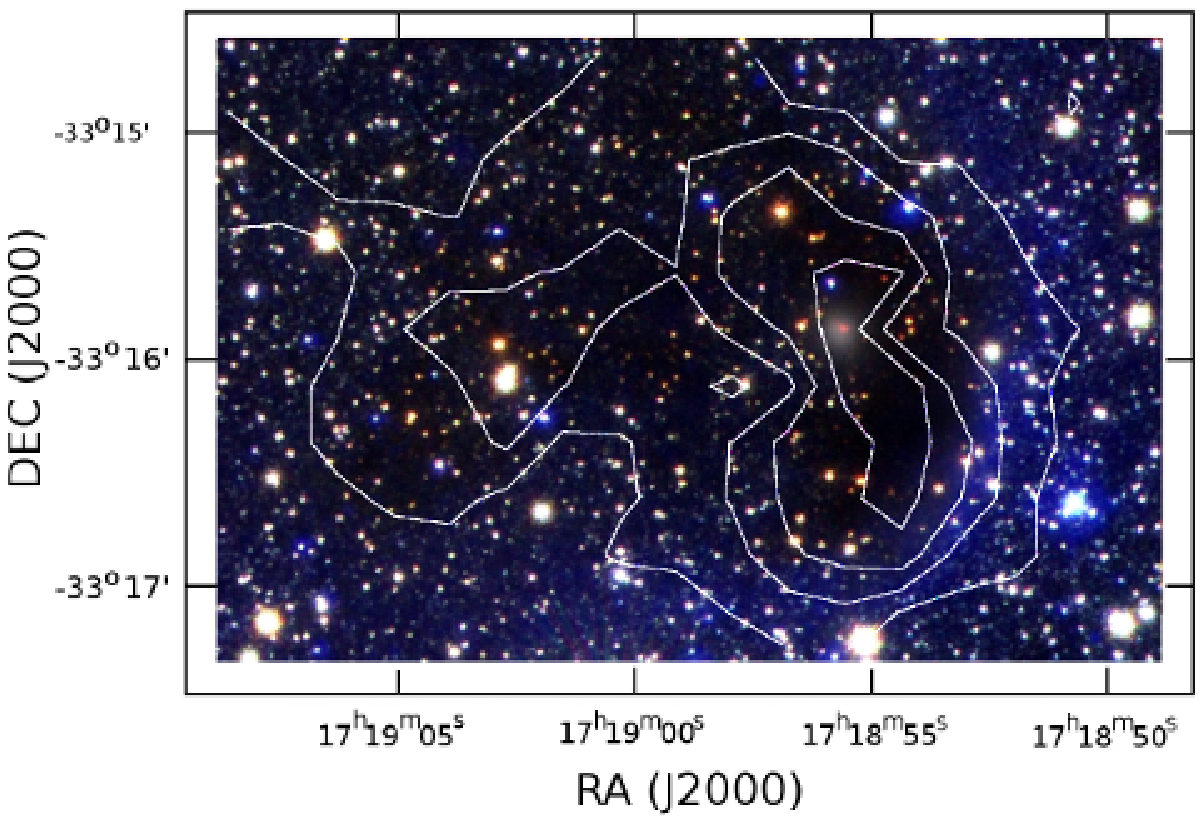}
   \caption{Noise estimate for the extinctions in Fig. \ref{fig:BR1_ext}.   The contours range from 2\umag \  to 5\umag in steps of  1\fm0.}
   \label{fig:BR1_ext_rms}
\end{figure}
  }

\newpage
  
  {\section{Additional figures} 

\begin{figure*}[h]
 \centering
 \includegraphics [width=18cm] {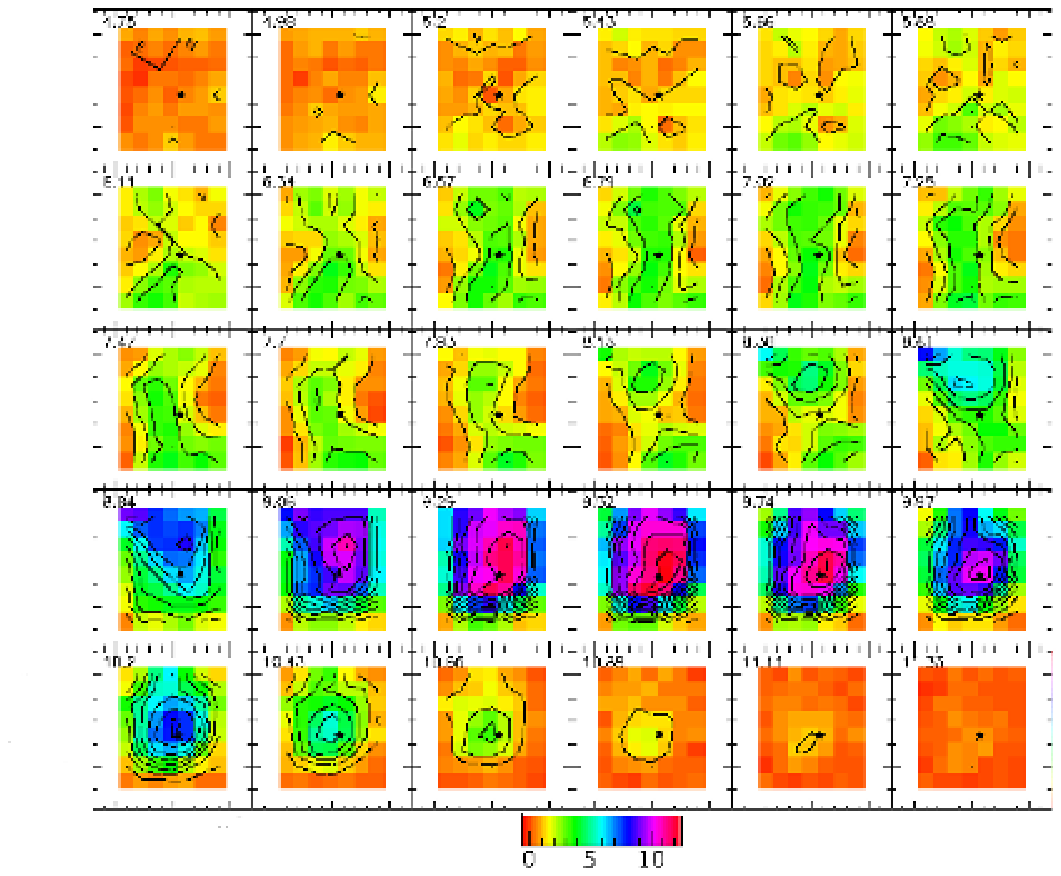}
    \caption{Channel map of CO (1--0) line emission in velocity bins
      of 0.23~\kmps. Pixel scale is 20\arcsec\ by 20\arcsec \ and the
      North-East orientation is indicated in the upper left panel. The
      position of BHR\,160\,IRS1 is indicated by an asterisk 
      in each panel.{ The lowest contour level and increment are 1.0\,\kkms. } }
 \label{fig:CO1-0_channel}
 \end{figure*}

\begin{figure*}[h]
 \centering
 \includegraphics [width=18cm] {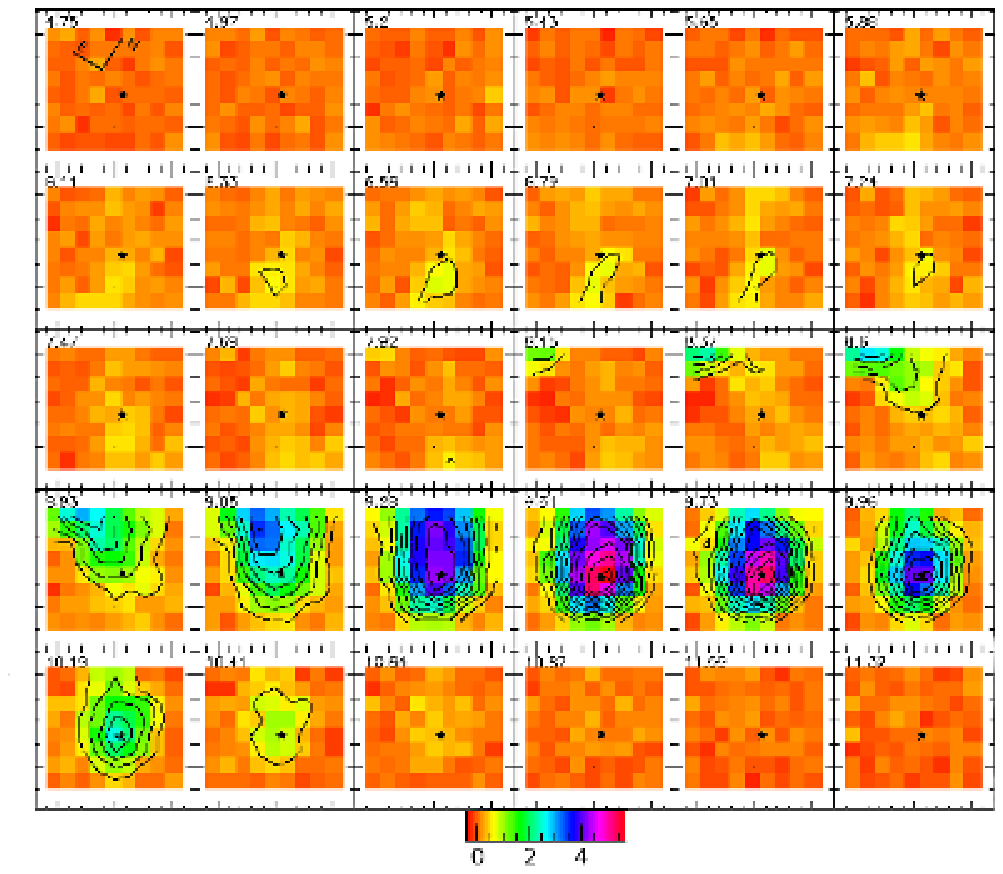}
    \caption{As Fig. \ref{fig:CO1-0_channel} for \thco (1--0). { The lowest contour level and increment are 0.5\,\kkms. } }
 \label{fig:13CO1-0_channel}
 \end{figure*}

\begin{figure*}[h]
 \centering
 \includegraphics [width=18cm] {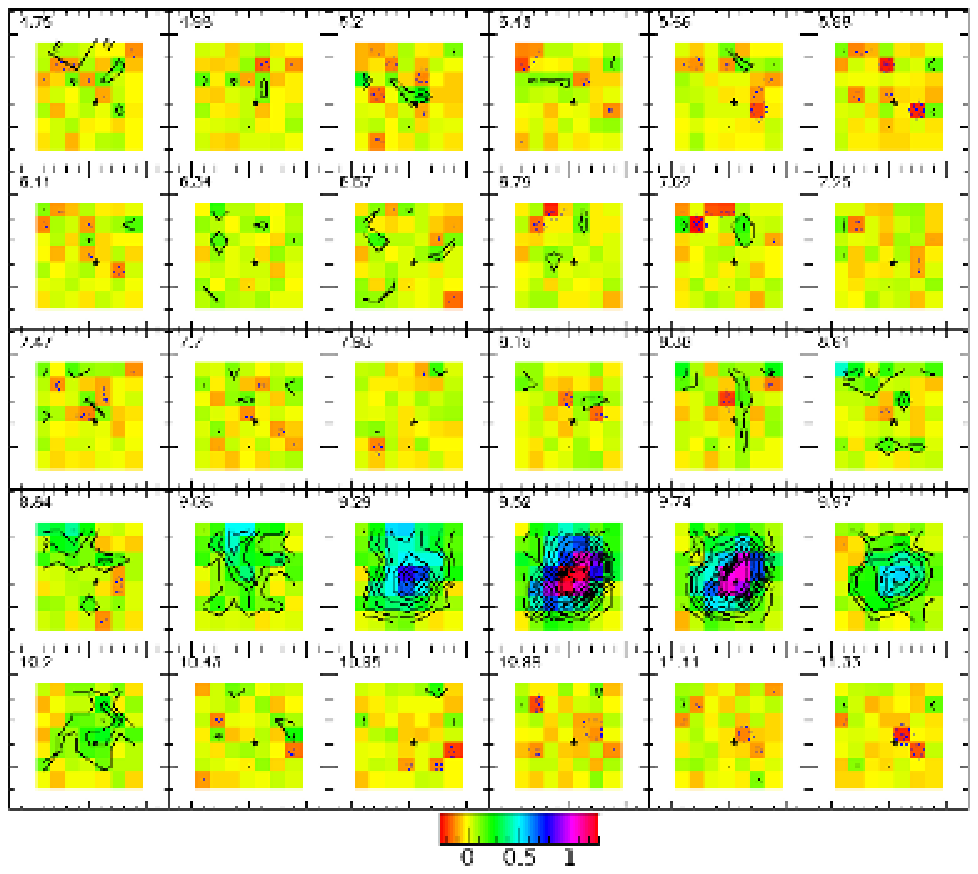}
    \caption{As Fig. \ref{fig:CO1-0_channel} for \ceo (1--0). { The lowest contour level and increment are  0.1\,\kkms. } }
 \label{fig:C18O1-0_channel}
 \end{figure*}

\begin{figure*}[h]
 \centering
 \includegraphics [width=18cm] {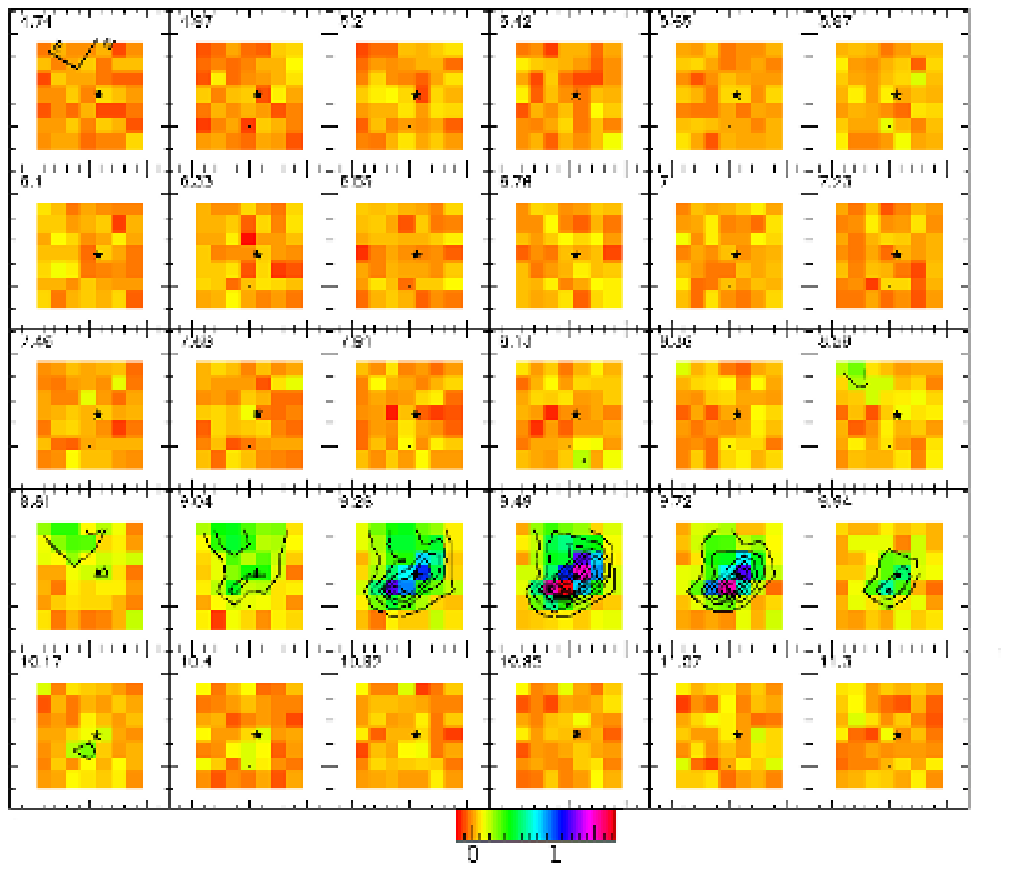}
    \caption{As Fig. \ref{fig:CO1-0_channel} for \ceo (2-1). { The lowest contour level and increment are 0.2\,\kkms. } } 
 \label{fig:C18O2-1_channel}
 \end{figure*}

\begin{figure*}[h]
 \centering
 \includegraphics [width=18cm] {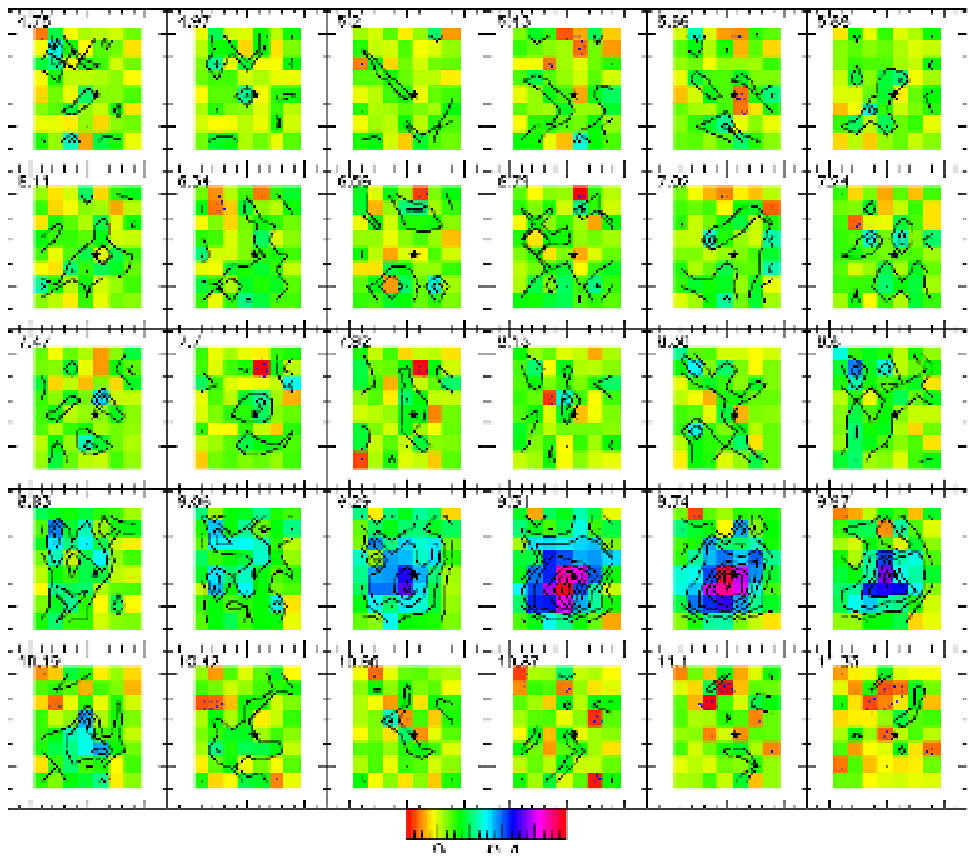}
    \caption{As Fig. \ref{fig:CO1-0_channel} for CS (2--1). { The lowest contour level and increment are 0.1\,\kkms. } } 
 \label{fig:CS2-1_channel}
 \end{figure*}

\begin{figure*}[h]
 \centering
 \includegraphics [width=18cm] {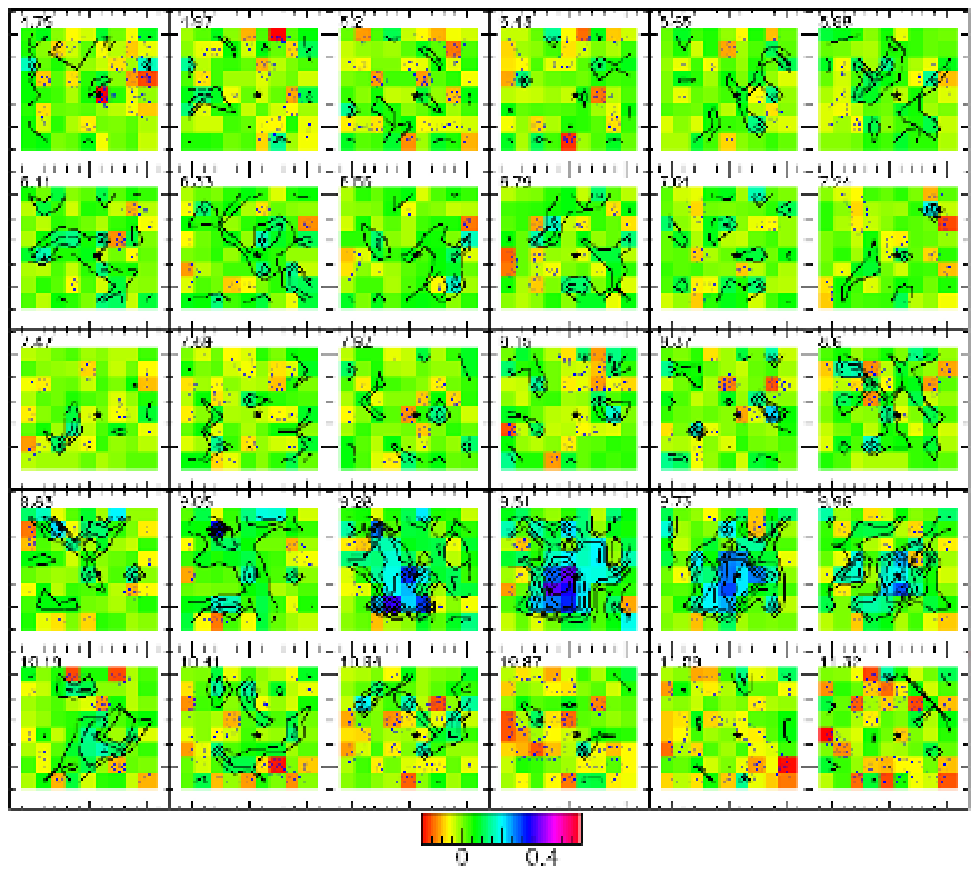}
    \caption{As Fig. \ref{fig:CO1-0_channel} for CS (3-2). { The lowest contour level and increment are  0.05\,\kkms. } }
 \label{fig:CS3-2_channel}
 \end{figure*}

\begin{figure}[h]
 \centering
 \includegraphics [width=8.8cm] {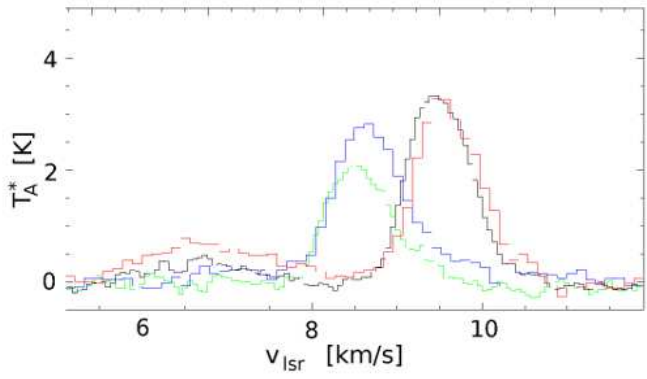}
    \caption{\thco (1--0) and (2--1) spectra in off-positions 0\arcsec,0\arcsec (red and black) and 135\arcsec,15\arcsec\ (blue and green). }
 \label{spectra-appendix}
 \end{figure}

\begin{figure*}[h]
 \centering
 \includegraphics [width=15cm]{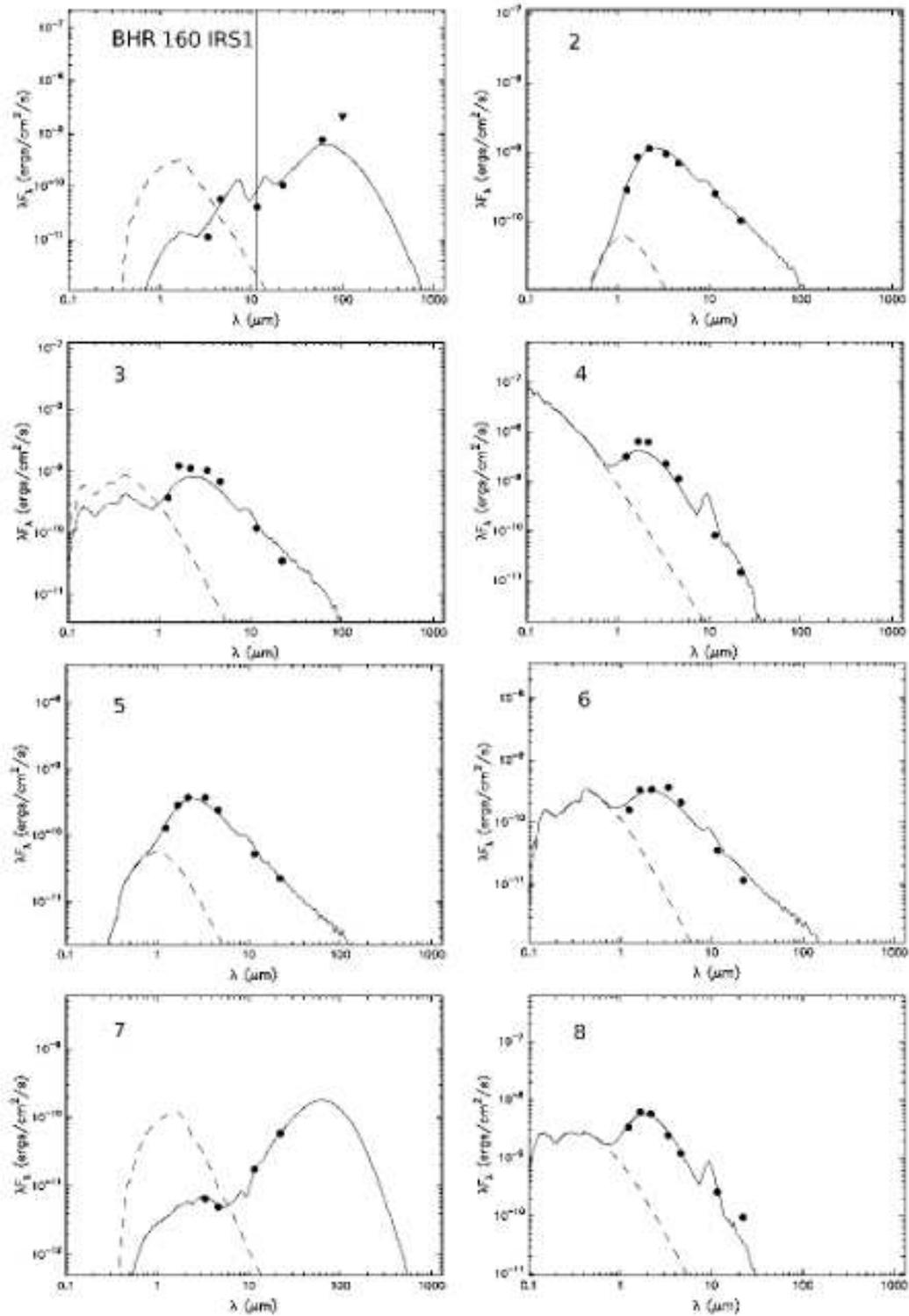}
 \caption{SED fits for the marked objects in Fig. \ref{fig:WISE}.
   The filled circles are the
observed fluxes and the triangle marks the  BHR\,160\,IRS1
IRAS 100\,$\mum$ flux  upper limit.
The dashed line indicates the SED of the { unreddened} stellar photosphere.}
 \label{fig:sed1}
\end{figure*}

}}

\end{appendix}

\end{document}